\documentclass{jfm}
\pdfoutput=1
\usepackage{graphicx}
\usepackage{subfigure}
\usepackage{natbib}
\usepackage{mathrsfs}

\ifCUPmtlplainloaded \else
  \checkfont{eurm10}
  \iffontfound
    \IfFileExists{upmath.sty}
      {\typeout{^^JFound AMS Euler Roman fonts on the system,
                   using the 'upmath' package.^^J}%
       \usepackage{upmath}}
      {\typeout{^^JFound AMS Euler Roman fonts on the system, but you
                   dont seem to have the}%
       \typeout{'upmath' package installed. JFM.cls can take advantage
                 of these fonts,^^Jif you use 'upmath' package.^^J}%
      }
  \else
  \fi
\fi

\ifCUPmtlplainloaded \else
  \checkfont{msam10}
  \iffontfound
    \IfFileExists{amssymb.sty}
      {\typeout{^^JFound AMS Symbol fonts on the system, using the
                'amssymb' package.^^J}%
       \usepackage{amssymb}%
       \let\le=\leqslant  
       \let\ge=\geqslant  
      }{}
  \fi
\fi

\ifCUPmtlplainloaded \else
  \IfFileExists{amsbsy.sty}
    {\typeout{^^JFound the 'amsbsy' package on the system, using it.^^J}%
     \usepackage{amsbsy}}
    {}
\fi

\newsavebox{\astrutbox}
\sbox{\astrutbox}{\rule[-5pt]{0pt}{20pt}}

\def\vb{v_\mathrm{ belt}}
\def\vn{v_\mathrm{ nozzle}}
\def\se{s_\mathrm{ end}}
\def\tm{\tau}
\def\tme{\tm_\mathrm{ end}}

\def\tmew{\tm_\mathrm{ end}(w)}

\def\be{\begin{equation}}
\def\ee{\end{equation}}

\def\vtw{v(\tm;w)}

\def\B{\mathrm{Re}}
\def\I{I}

\def\ves{{\mathbf e}_t}
\def\ven{{\mathbf e}_n}
\def\mcA{{\mathcal A}}
\def\lrho{\rho \mcA}
\def\lK{{\mathbf K'}}
\def\vr{{\mathbf r}}
\def\vg{{\mathbf g}}

\def\na{\alpha_\mathrm{ nozzle}}

\def\Acr{A^*}

\def\L{L}
\def\D{d_\mathrm{ nozzle}}
\def\xe{x_\mathrm{ end}}
\def\F{F}
\def\Dr{\mathrm{Dr}}

\def\Iv{I_\mathrm{vert}}
\def\Ix{I_\mathrm{conv}}
\def\Ic{I_\mathrm{conc}}

\def\vvtw{v_\mathrm{vert}(\tm;w)}
\def\vxtw{v_\mathrm{conv}(\tm;w)}
\def\vctw{v_\mathrm{conc}(\tm;w)}

\def\Vv{\mathscr{P}_\mathrm{vert}}
\def\Vx{\mathscr{P}_\mathrm{conv}}
\def\Vc{\mathscr{P}_\mathrm{conc}}
\def\V{\mathscr{P}}

\def\Cxe{\mathcal{C}(\xe,\vb)}
\def\msR{\mathscr{R}}
\def\msD{\mathscr{D}}

\title[Three viscous jet flow regimes]{Three flow regimes of viscous jet falling onto a moving surface}

\author[A. Hlod, A.C.T. Aarts, A.A.F. van de Ven and  M.A. Peletier]%
{A. Hlod$^1$%
  \thanks{Present address: Dept. of Mathematics and Computer Science Technische Universiteit Eindhoven PO Box 513 5600 MB  Eindhoven The Netherlands }, A.C.T. Aarts$\,^1$, A.A.F. van de Ven$\,^1$ \and  M.A. Peletier$\,^1$
}

\affiliation{$^1$Center for Analysis, Scientific computing and Applications, Eindhoven University of Technology, Eindhoven, The Netherlands.}

\date{\today}

\begin{document}

\maketitle

\begin{abstract}
A stationary viscous jet falling from an oriented nozzle onto a moving surface is studied, both theoretically and experimentally. We distinguish three flow regimes and classify them by the convexity of the jet shape (concave, vertical and convex). The fluid is modeled as a Newtonian fluid, and the model for the flow  includes viscous effects, inertia and gravity. By studying the characteristics of the conservation of momentum for a dynamic jet, the boundary conditions for each flow regime are derived, and the flow regimes are characterized in terms of the process and material parameters.  The model is solved by a transformation into an algebraic equation. We make a comparison between the model and experiments, and obtain qualitative agreement.
\end{abstract}
\section{Introduction} \label{sec:intr}
%
%
During the fall under gravity of a viscous jet from an oriented nozzle onto a moving surface, called the belt, three flow regimes can be distinguished. The flow regimes are characterized by the jet shape and depend on the process parameters such as dynamic viscosity of the Newtonian fluid, flow velocity at the nozzle, belt velocity, and falling height.

In the first flow regime, apart from a boundary layer at the belt, the jet shape is concave and aligned with the nozzle orientation, and resembles a ballistic trajectory;  see Figures~\ref{fig:Experiment1}~and~\ref{fig:Experiment2}.  We call this flow regime \emph{concave} and the associated jet a \emph{concave jet}. The concave jet occurs for large flow velocity at the nozzle and small viscosity.

The jet in the second flow regime has a straight, vertical shape, apart from boundary layers at the nozzle and at the belt; see Figure~\ref{fig:Experiment3}. We call this flow regime \emph{vertical}, and  the associated jet a \emph{vertical jet}. The vertical jet happens for large falling heights, large viscosity, and small flow velocity at the nozzle.

In the third flow regime,  apart from a boundary layer at the nozzle, the jet shape is convex, and the jet touches the belt tangentially; see Figure~\ref{fig:Experiment4}. This flow regime we call \emph{convex}, and the associated jet a \emph{convex jet}. The convex jet occurs for high fluid viscosity, large belt velocity, small velocity at the nozzle, and small falling height.

This paper is a continuation and a generalization of our previous work on the convex jet (\cite{Hlod1}).
%

The fall of viscous jets or sheets from a nozzle oriented vertically down onto a \emph{fixed} surface has been widely studied. Here one can observe unstable behavior; see \cite{Taylor}, \cite{Scorobogatiy}, \cite{YarinTch}, \cite{Ribe1}, \cite{Ribe2}, \cite{Cruickshank} and \cite{TchYarinRad}. Vertically falling viscous jets have been studied in \cite{Clarke66}, \cite{Clarke68}, \cite{Adachi} and \cite{Sauter}. Experimental investigations of steady and unsteady flows of a viscous jet falling under gravity onto a \emph{moving} surface from a vertical nozzle were presented in \cite{Lister} and \cite{morris:066218}. In \cite{Lister} and \cite{Lister2} the steady flow is modeled and the parameter region of the steady flow is determined in terms of the falling height and the surface velocity.
%

However, the previous publications make no distinction between concave and vertical flows, and because the nozzle is oriented vertically down the concave flow is not recognized as a separate regime. In this paper, we fully describe all three flow regimes. To describe the jet we use a model which includes effects of inertia, viscosity and gravity, but neglects surface tension, bending stiffness and air drag. The fluid is considered to be incompressible, Newtonian, and temperature effects are neglected. We allow the nozzle orientation to vary between horizontal and vertically down. By studying the characteristics of the equation of momentum conservation, we determine the parameter regions for each flow regime. Consideration of the characteristics as being the directions of information propagation explains why and when each of the three flow regimes occurs and gives the correct boundary conditions for each flow regime. To validate our theoretical results we perform experiments of the jet falling from the oriented nozzle onto the moving belt. We find a qualitative agreement between the experimentally observed and the theoretical values of the positions of the touchdown points for different belt velocities. The model presented in this paper can also be used to describe the fall of viscous sheets onto a moving surface.
%

The structure of the paper is as follows: In Section~\ref{sec:experiments}, we describe the experiments of the fall of the viscous jet onto a moving belt, and present the experimental results. In Section~\ref{sec:Modelling} the model equations are derived and simplified to a first-order differential equation on unknown domain. The analysis of the characteristics of the conservation of momentum equation for dynamic jets in order to derive correct boundary conditions is given in Section~\ref{sec:ModelBC}. In Section~\ref{sec:ResultsFromModel} we present some results from the model, and in Section~\ref{sec:Comparison} we compare them with experiments. The characteristic features of the three flow regimes are summarized in \ref{sec:SummaryThreeFR}, and some conclusions are made in Section~\ref{sec:Conclusions}.
\section{Experiments} \label{sec:experiments}
In this section we describe experiments of the fall of a thin jet of a Newtonian fluid onto a moving belt. We focus on the shape of the jet between the nozzle and the belt. We describe the experimental setup, report our observations and present some conclusions from the experiments.
\subsection{Experimental method}\label{sec:expMethod}
\begin{figure}
\center\includegraphics[width=0.7\textwidth]{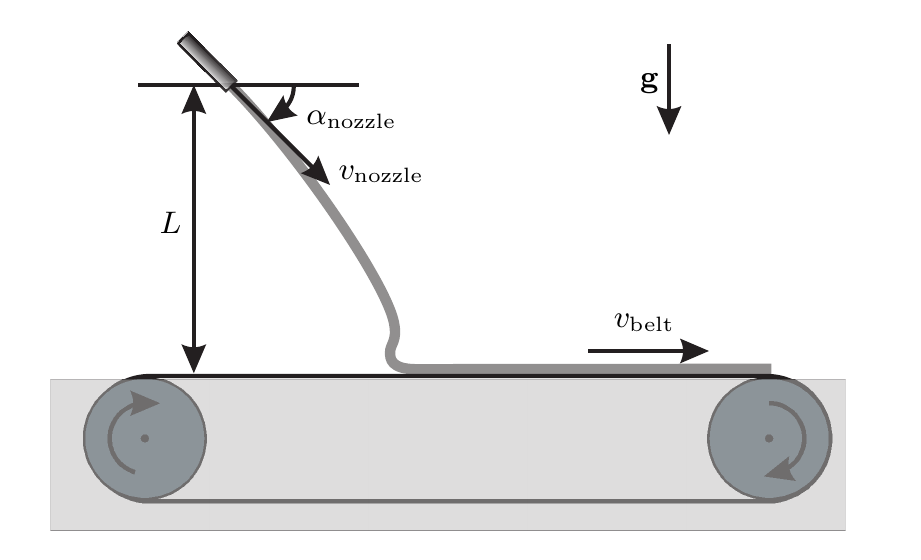}
   \caption{The scheme of the experimental setup for a jet falling onto a moving belt.}
   \label{fig:ExperSetup}
\end{figure}

A viscous fluid, polybutene Indopol H-100, is pumped to a nozzle and allowed to fall from the nozzle onto a moving belt; see Figure~\ref{fig:ExperSetup}. The belt is wrapped around two horizontal cylinders at the same height. The left cylinder is connected to an electric motor, to move the horizontal belt from the left to the right with a constant speed.
\begin{table}
   \begin{center}
     \begin{minipage}{8cm}
        \begin{tabular}{{|l|l|l|l|}}
          \hline
            Parameter name & & Value & Unit\\
          \hline
          belt velocity & $\vb$ & 0\ -\ 5 & m/s \\
          flow velocity at nozzle & $\vn$ & 0.4\ -\ 1.2 & m/s \\
          distance between belt and nozzle & $\L$ & 0.01\ -\ 0.07 & m \\
          nozzle orientation\footnote{The angle between the nozzle orientation and the horizontal direction,
          positive for downwards-pointing nozzle.}
           & $\na$ &   -9\ -\ 38$^\circ$ & \\
          kinematic viscosity of fluid & $\nu$ & 0.047 & $\mathrm{m^2/s}$ \\
          fluid density & $\rho$ & 880 & $\mathrm{kg/m^3}$\\
          nozzle diameter & $\D$ & 1 or 0.4 & mm \\
          \hline
       \end{tabular}
     \end{minipage}
   \end{center}
  \caption{Values of the experimental parameters} \label{tb:ExpValues}
\end{table}

The nozzle is placed above the belt. The nozzle - belt distance and the belt and the nozzle orientation can be varied. A screw pump producing a constant flow rate is connected to the nozzle. The flow rate was measured by weighing the fluid collected from the nozzle during 30 s. In the experiments, two different nozzles were used, with diameters of $1\,\mathrm{mm}$ and $0.4\,\mathrm{ mm}$.

The experimental setup allows us to change the nozzle position and orientation, belt velocity, and flow velocity from the nozzle. For all experiments the same fluid is used. The fluid is considered to be Newtonian. No nonlinear effects such as die swell near the nozzle were observed. The values of the experimental parameters are given in Table~\ref{tb:ExpValues}.

\subsection{Experimental results}\label{sec:expResults}
\begin{figure}
\centering
 \subfigure[$\vb=0.093\, \mathrm{m/s}$ ]
  {\label{fig:Experiment1}\includegraphics[width=0.4\textwidth, height=0.42\textwidth]{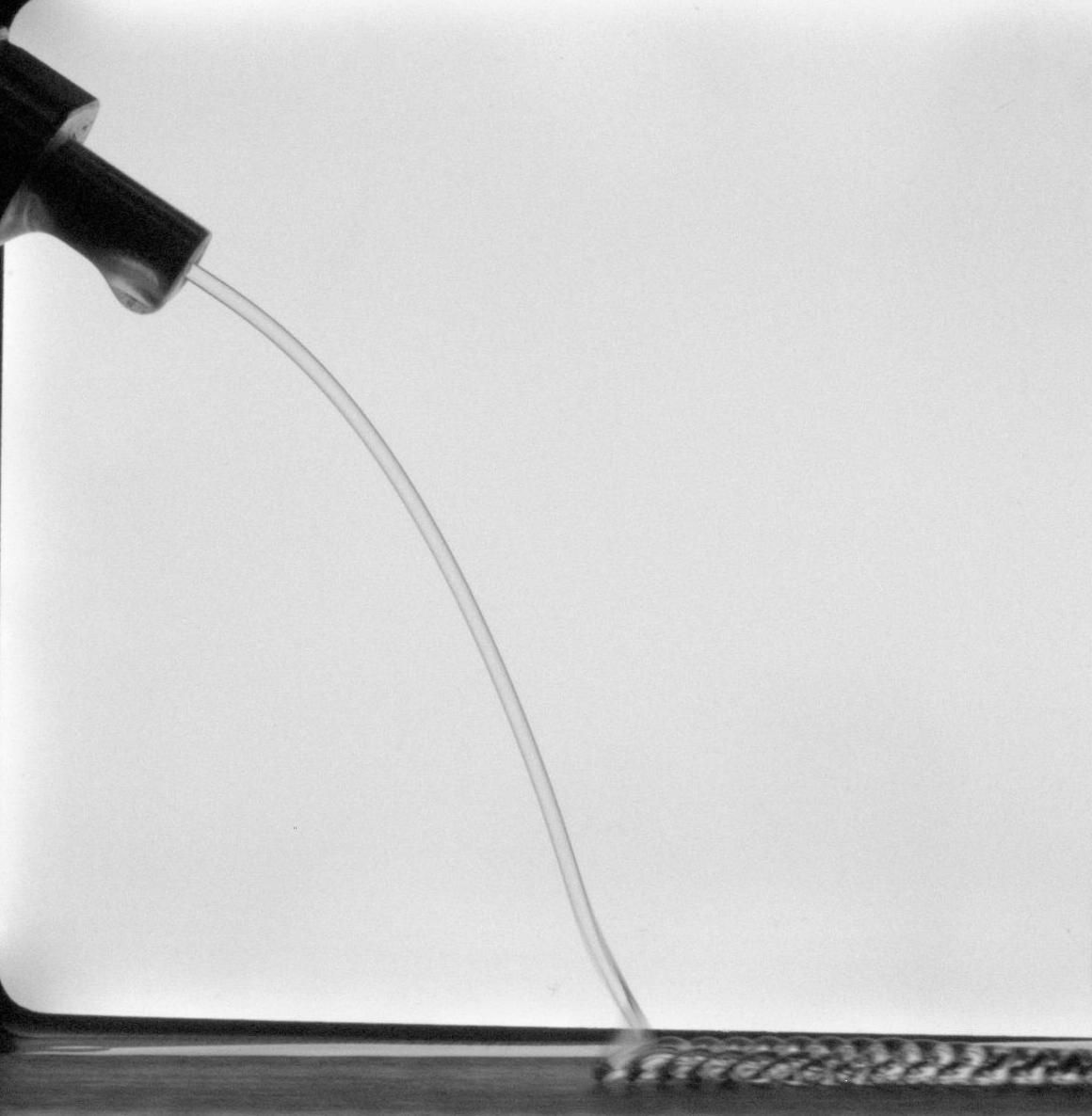}}\
 \subfigure[$\vb=0.750\, \mathrm{m/s}$ ]
  {\label{fig:Experiment2}\includegraphics[width=0.4\textwidth, height=0.42\textwidth]{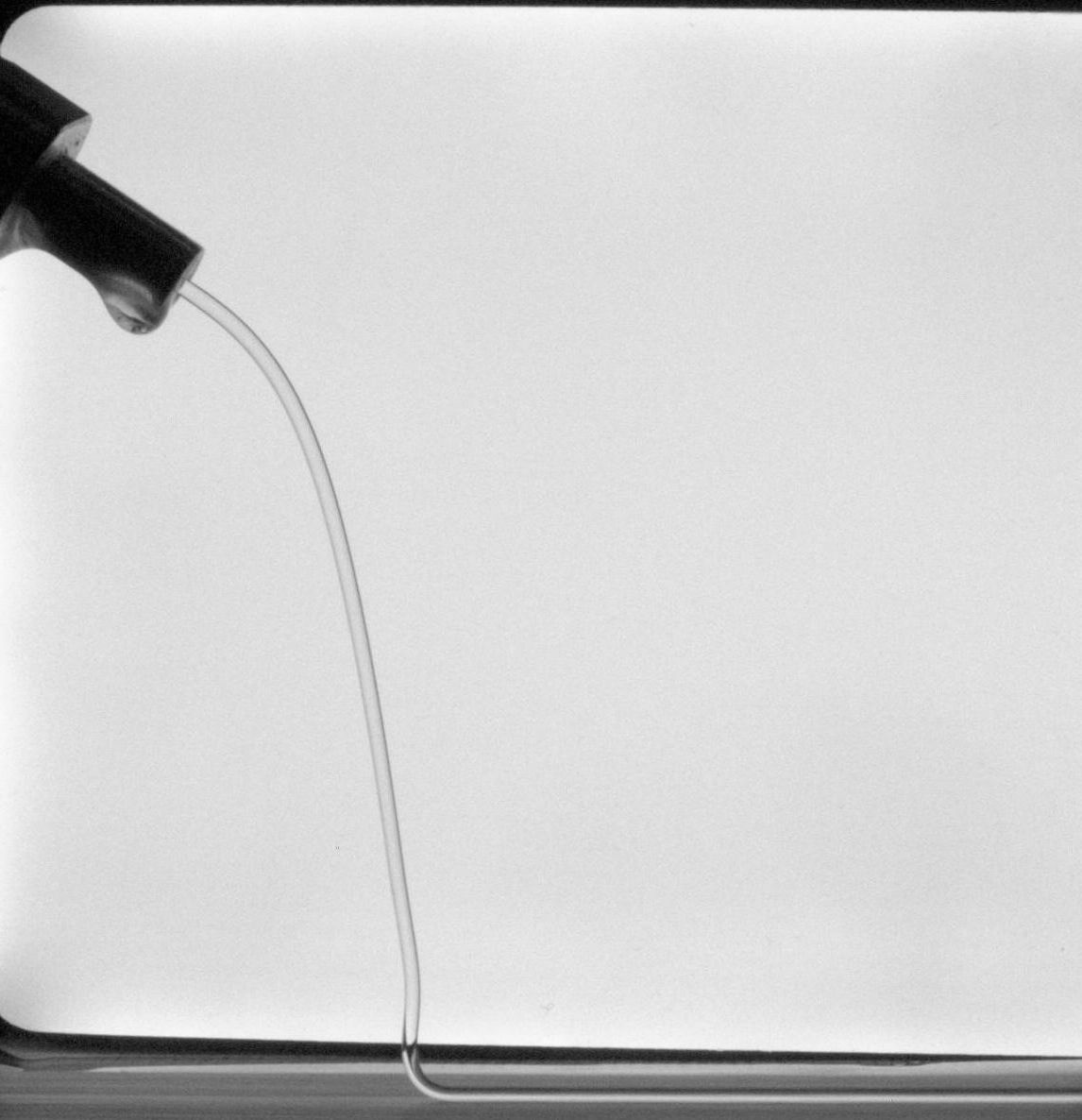}}\\
  \subfigure[$\vb=1.708\, \mathrm{m/s}$ ]
  {\label{fig:Experiment3}\includegraphics[width=0.4\textwidth, height=0.42\textwidth]{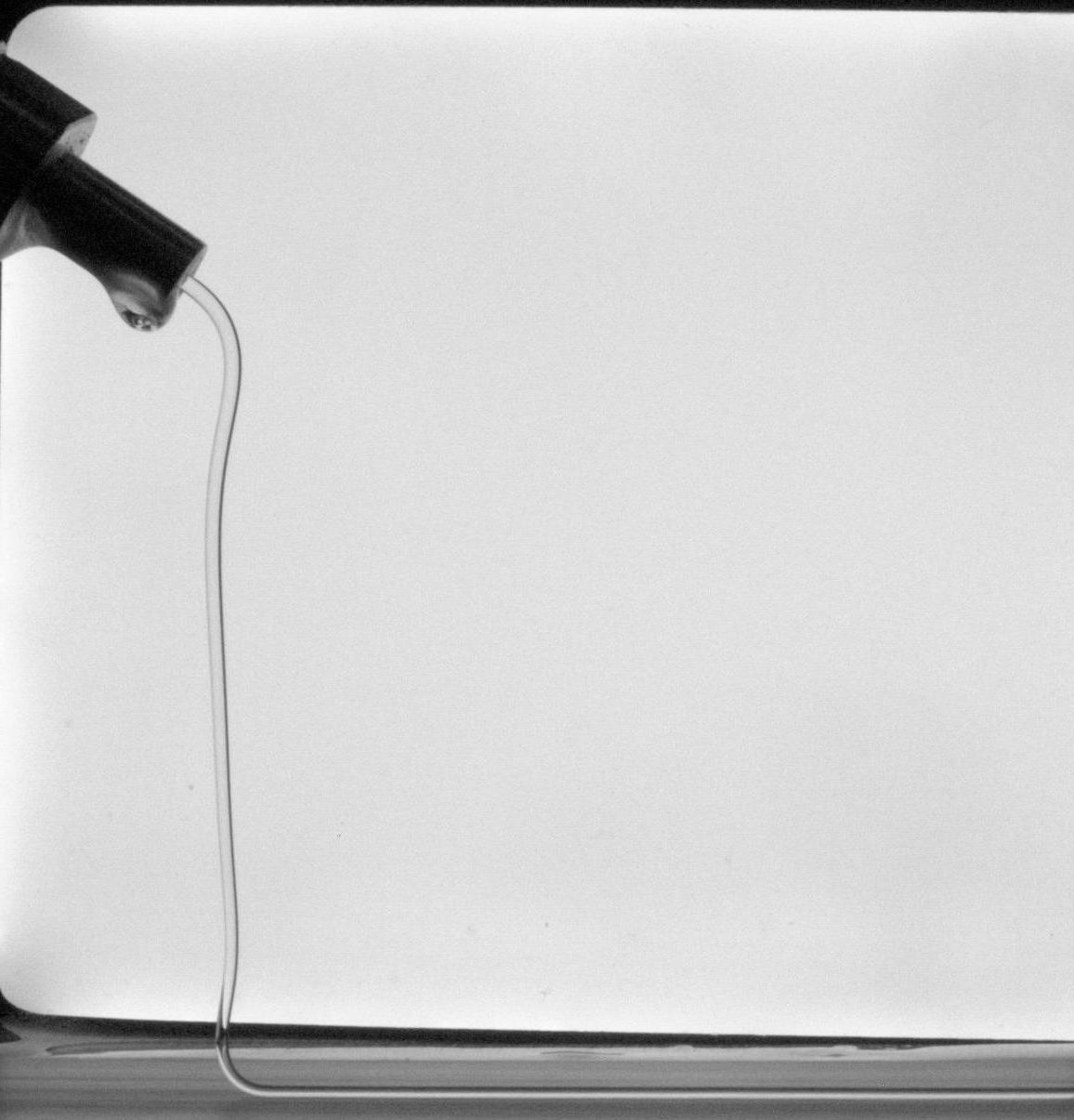}}\
 \subfigure[$\vb=3.200\, \mathrm{m/s}$ ]
  {\label{fig:Experiment4}\includegraphics[width=0.4\textwidth, height=0.42\textwidth]{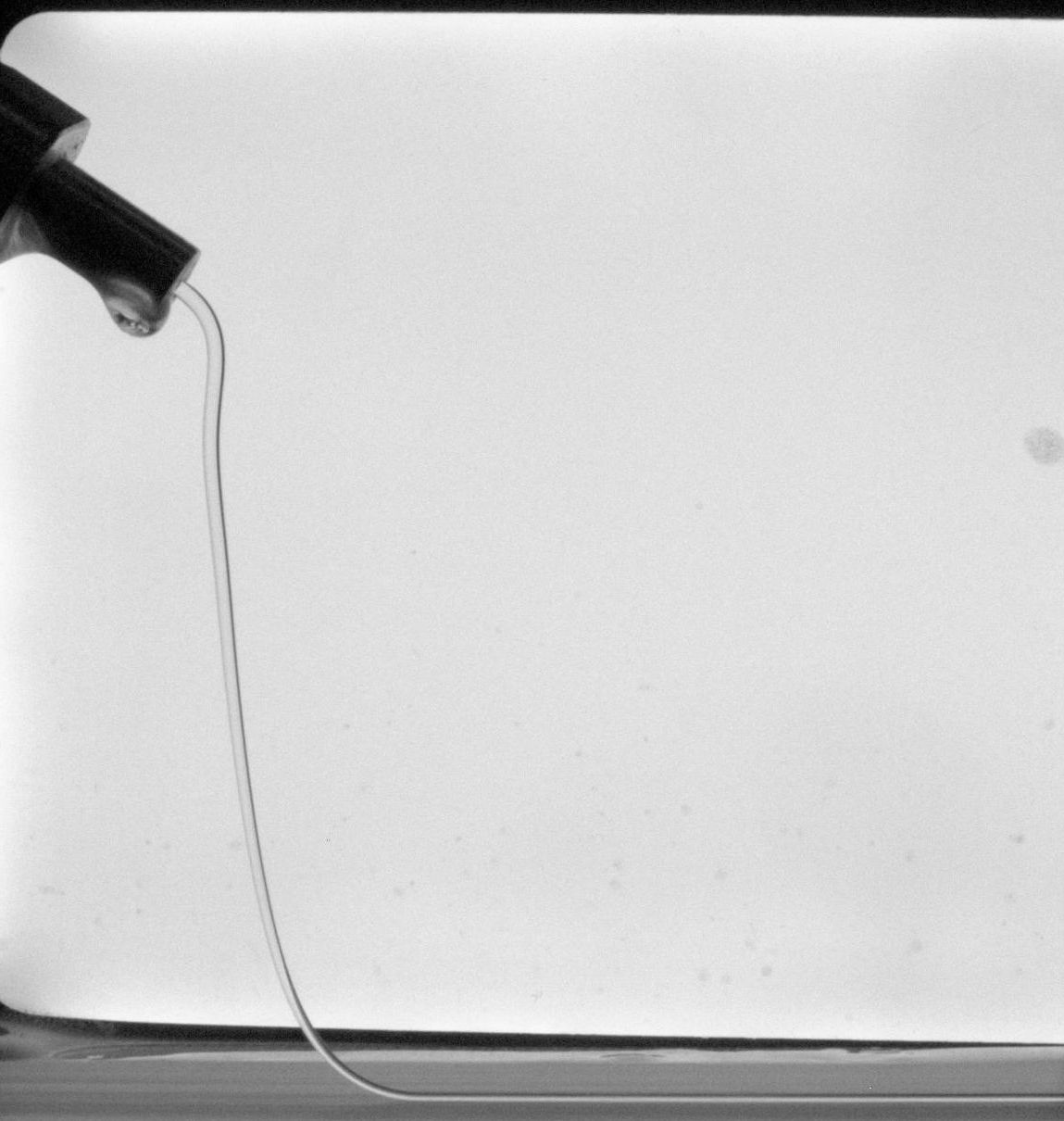}}\\
   \caption{Jet shapes for the set of experiments with $\D=1\,\mathrm{mm}$, $\vn=1.061\,\mathrm{m/s}$, $\L=0.054\,\mathrm{m}$ and $\na=37.3^\circ$. The belt moves from the left to the right. The jet shape changes from concave (Figures \ref{fig:Experiment1} and \ref{fig:Experiment2}) to vertical (Figure \ref{fig:Experiment3}), and from vertical to convex (Figure \ref{fig:Experiment4}), while $\vb$ is increased.}
   \label{fig:Experiments}
\end{figure}
First, we describe a typical sequence of experiments. We start with  $\vb$ close to zero and make sure that $\L$ and $\vn$ are chosen such that the shape of the jet is concave, resembling a ballistic trajectory; see Figure~\ref{fig:Experiment1}. To obtain the concave jet shape the nozzle should not point down vertically, and therefore we put $\na<\pi/2$. Next, we gradually increase $\vb$ and study the evolution of the jet shape.

For small $\vb$ the jet shape is concave with an unstationary region near the belt; see Figure~\ref{fig:Experiment1}. By increasing $\vb$, we observe that the unstationary region near the belt transforms into a stable bending region where the jet bends to the horizontal belt direction; see Figure~\ref{fig:Experiment2}. The jet shape in this region resembles the backward-pointing heel, reported for the vertically falling jet in \cite{Lister}.

When we increase $\vb$ further, the jet shape approaches the vertical direction. Note that this direction of change is often considered counterintuitive.  In this case, the contact point with the belt approaches the vertical projection of the nozzle position. As a result, for $\vb$ large enough, the main part of the jet between the belt and the nozzle is purely vertical; see Figure~\ref{fig:Experiment3}. The bending region near the belt remains, and a new bending region near the nozzle appears. Near the nozzle the jet bends from the nozzle orientation to the vertical direction.

Further increase in $\vb$ results in the disappearing of the bending region near the belt. The jet shape becomes convex everywhere, except for a bending region near the nozzle; see Figure~\ref{fig:Experiment4}. The touchdown point moves away from the nozzle in the direction of the belt motion as $\vb$ increases.

Summarizing the results of the experiments, we observe a concave jet shape for small $\vb$, except for a small bending or unstable region near the belt. With increasing $\vb$ the jet shape becomes vertical, except for small bending regions near the nozzle and the belt. Further increase of $\vb$ leads to a convex jet shape, except for a small bending region near the nozzle. This gives a characterization of the jet flow by its shape, i.e. concave, vertical and convex.
\begin{figure}
\centering
 \includegraphics[width=0.8\textwidth]{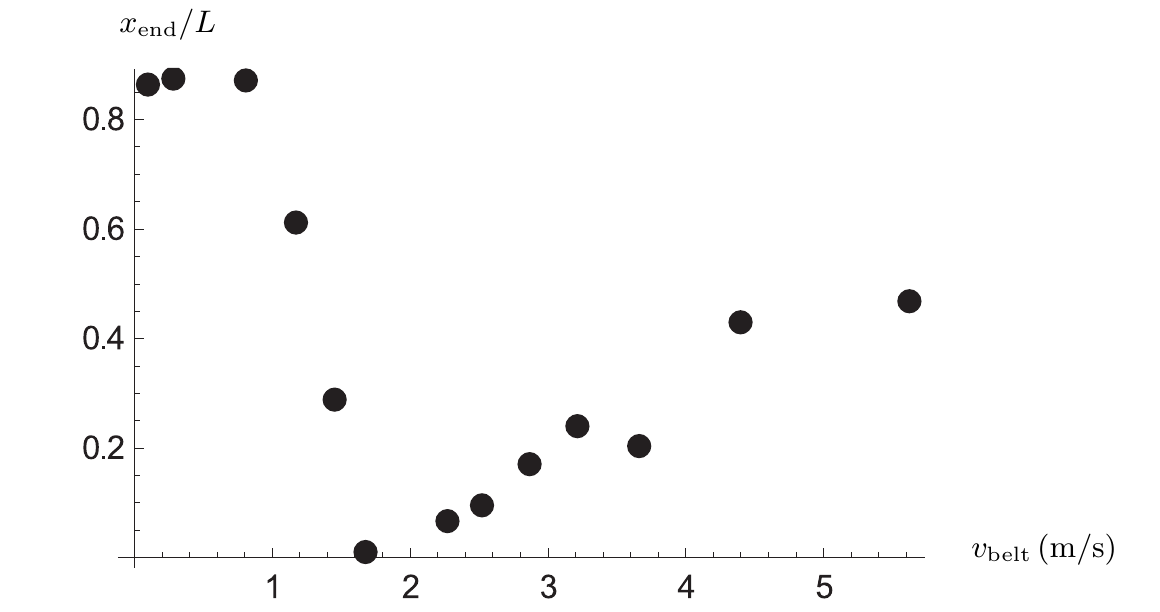}
  \caption{Experimentally obtained positions of the touchdown point $\xe$ for different $\vb$ indicated by $\bullet$. The other parameters are $\D=0.4\,\mathrm{mm}$,  $\vn=1.147\,\mathrm{m/s}$, $\L=4.1\,\mathrm{cm}$ and $\na=5^\circ$.}
   \label{fig:xendpos}
\end{figure}
\begin{figure}
\centering
 \includegraphics[width=0.8\textwidth]{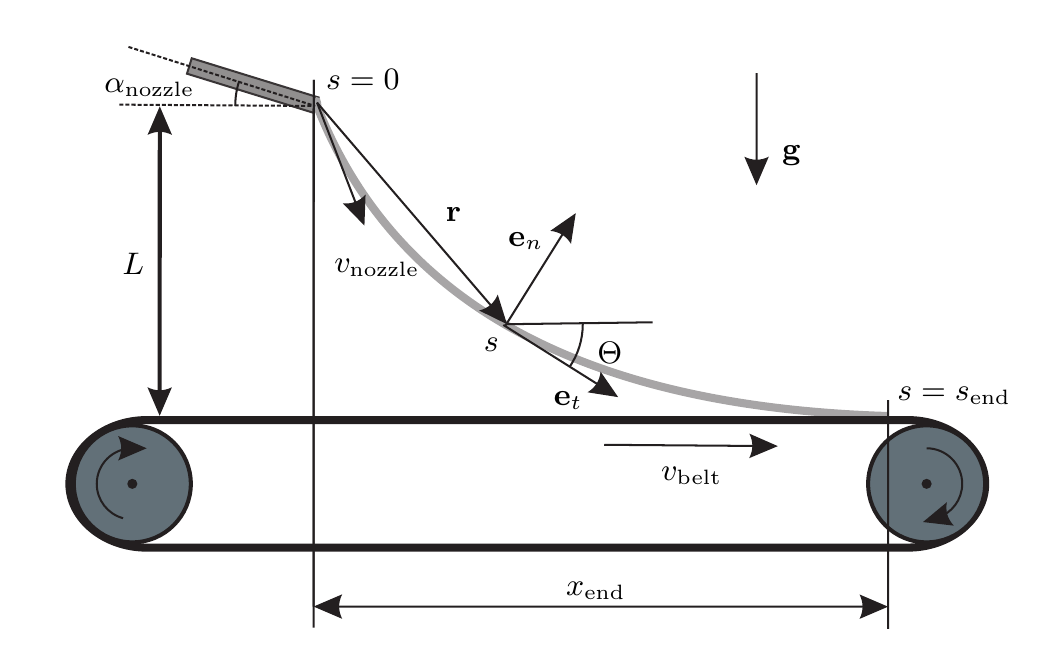}
  \caption{Model of a fall of a viscous jet onto a moving surface.}
   \label{fig:model}
\end{figure}

A convenient way to quantitatively study the jet is to look at the horizontal position $\xe$ (see Figure \ref{fig:model}) of the touchdown point at the belt, relative to the nozzle position. For the concave jet with $\vb$ small the contact point is away from the nozzle in the direction of the nozzle. With increasing $\vb$, the touchdown point moves first towards the nozzle position ($\xe$ decreases) until the jet becomes vertical ($\xe=0$), stays vertical for some time and then moves away from the nozzle position in the direction of the belt motion ($\xe$ increases). Figure~\ref{fig:xendpos} suggest that the first five dots are in the concave flow regime, the sixth, with $\xe=0$, represents a vertical flow, and the remaining ones (7th and higher) are in the convex flow regime.

\begin{figure}
 \center\includegraphics[width=0.6\textwidth]{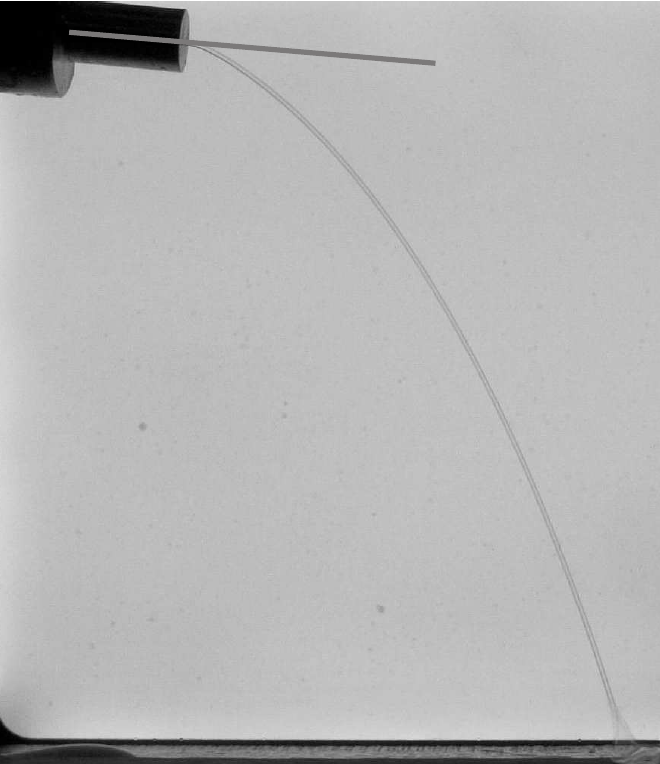}
   \caption{Experiment with $\D=0.4\,\mathrm{mm}$, $\vn=1.188\,\mathrm{m/s}$, $\vb=0.1\,\mathrm{m/s}$, $\L=0.068\,\mathrm{m}$ and $\na=5^\circ$. The jet has a concave shape, which is not aligned with the nozzle orientation.}
   \label{fig:ExpNozzleAngle}
\end{figure}
For the concave jet in Figures~\ref{fig:Experiment1}~and~\ref{fig:Experiment2}, where the nozzle diameter is 1~mm, the jet near the nozzle is aligned with the nozzle orientation. When the nozzle diameter is 0.4~mm, the jet is not aligned with the nozzle orientation. This can be seen in Figure~\ref{fig:ExpNozzleAngle}.

In the next section we present a model that characterizes the flow type and, among other things, predicts the relation between $\xe$ and $\vb$.
\section{Modeling and analysis} \label{sec:Modelling}
In this section we present our model of the fall of a viscous jet onto a moving belt. To model the flow we use a thin-jet approximation and include effects of inertia, viscous tension and gravity. We assume the fluid to be incompressible, isothermal, and Newtonian. We neglect surface tension, bending stiffness, and air drag. Therefore possible bending or buckling regions at the nozzle or at the belt are disregarded. The jet is described by the equations of conservation of mass and momentum. First we formulate the equations for the dynamic jet, which are used in Section~\ref{sec:ModelBC} to justify or choice of boundary conditions. Next, we partly solve the stationary jet equations and make an analysis showing that the jet can have only three possible shapes: concave, vertical and convex. Finally, we reformulate the problem by deriving an equivalent algebraic equation which is convenient for further analysis.

The jet is modeled as a curve in the $x,z$-plane of unknown length $\se$,; see Figure~\ref{fig:model}. The curve is parameterized by its arc length $s$, with the origin $s=0$ at the nozzle and $s=\se$ at the touchdown point at the belt. The position of a certain point $s$ of the thin jet at time $t$ is described by its position vector $\vr=\vr(s,t)$ with respect to the origin $0$, which is chosen at the nozzle point.

A local coordinate system in a point $s$ having as basis the tangent and normal vectors $\ves,\,\ven$, is constructed at each point of the jet. The angle between the tangent vector and horizontal direction is $\Theta$. The horizontal distance between the nozzle and the touchdown point at the belt is $\xe$. The flow velocity in a point $s$ of the jet is $v=v(s,t)$. The jet at the touchdown point has the same velocity as the belt $\vb$, and the flow velocity at the nozzle is $\vn$.

The system of equations describing a thin dynamical jet in two dimensions can be found in variety of publications i.e. \cite{Roos}, \cite{Yarin}, and \cite{YarinEntonovDJ}. It consists of the laws of conservation of mass and momentum
\be
 \mcA_t+(\mcA v)_s=0, \label{eq:consMass}
\ee
\be
 \lrho (\vr_{tt}+ \vr_{s}(v_t+v v_s)+v^2 \vr_{ss}+2 v \vr_{st})=P_s \vr_s+P \vr_{ss}+\lK, \label{eq:consMom}
\ee
respectively, where $\mcA=\mcA(s,t)$  is the cross-sectional area,  $P=P(s,t)$ is the longitudinal force, and $\lK=\lK(s,t)$ is the external force per unit of length of the jet. The longitudinal force $P$ is given by a constitutive law, and in the case of a Newtonian viscous fluid it is equal to
\be
    P=3 \nu \rho v_s \mcA. \label{eq:constLawFluid}
\ee
 Finally for $\lK$ we take
\be
\lK= \lrho \vg, \label{eq:BodyForce}
\ee
the gravity force per unit of length of the jet (external air drag is neglected).

The stationary versions of the equations (\ref{eq:consMom})-(\ref{eq:consMass}) together with (\ref{eq:constLawFluid})-(\ref{eq:BodyForce}), and the condition for $s$ as the arc length are
\begin{eqnarray}
     \mcA(\vr_{s} v v_s+ v^2 \vr_{ss})=3 \nu (v_s \mcA \vr_s)_s+\mcA \vg, \label{eq:stjet_cons_mom1}\\
     (\mcA v)_s=0,  \label{eq:stjet_cons_mass1} \\
     | \vr_s|=1. \label{eq:stjet_ArcLength1}
\end{eqnarray}
Thus, we have three differential equations, (\ref{eq:stjet_cons_mom1})-(\ref{eq:stjet_ArcLength1}), for the unknowns $\vr$, $v$ and $\mcA$. Next we describe the boundary conditions.

For  the velocity $v$ we prescribe two boundary conditions: at $s=0$, the flow velocity at the nozzle is
\be
    v(0)=\vn, \label{eq:bcVAtNozzle}
\ee
while at $s=\se$ the jet sticks to the belt, so
\be
    v(\se)=\vb. \label{eq:bcVAtBelt}
\ee
The boundary condition for $\mcA$ follows form the known cross-sectional area of the nozzle as
\be
    \mcA(0)=\frac{\pi}{4} \D^2. \label{eq:bcLRhoAtNozzle}
\ee
The fixed vertical distance between the nozzle and the belt gives the additional constraint
\be
    \int_0^{\se} \sin\Theta(s)\,ds=\L. \label{eq:YConstr}
\ee
To make the system (\ref{eq:stjet_cons_mom1})-(\ref{eq:YConstr}) complete we need two boundary conditions for $\vr$. Since the position $\vr$ is with respect to the fixed nozzle, we have one boundary condition for $\vr$
\be
    \vr(0,t)=\mathbf{0}. \label{eq:bcVRAtNozzle}
\ee
The second boundary condition is chosen later in this section and our choice is justified in Section~\ref{sec:ModelBC}.
By integrating  (\ref{eq:stjet_cons_mass1}), using (\ref{eq:bcVAtNozzle}) and (\ref{eq:bcLRhoAtNozzle}), we find that
$$
    \mcA(s)=\frac{\F}{v(s) \rho},
$$
where the mass flux is given by  $\F=\rho \vn \pi \D^2/4$. We eliminate $\mcA$ from (\ref{eq:stjet_cons_mom1}) to obtain
\be
    \vr_{s} v_s+v \vr_{ss}=3 \nu (\vr_s v_s/v)_s+ \vg/v. \label{eq:stjet_cons_mom2}
\ee
Next we introduce a new variable $\xi$ by
\be
    \xi=v-3 \nu\frac{v_s}{v}, \label{eq:Defxi}
\ee
which stands for the scaled momentum transfer through a jet cross-section and plays a crucial role in our further analysis. By use of $\xi$ we write (\ref{eq:stjet_cons_mom2}) as
\be
     (\xi \vr_s)_s=\frac{\vg}{v}. \label{eq:stjet_cons_mom3}
\ee
Using $\ves=\vr_s$, and $(\ves)_s=-\Theta_s \ven$, we can write (\ref{eq:stjet_cons_mom3}) in components as
\be
   \xi_s = \frac{g \sin(\Theta)}{v}, \label{eq:GSTheta1}
\ee
 and
\be
   \Theta_s = \frac{ g\cos (\Theta)}{\xi v}. \label{eq:GSTheta2}
\ee
Equation (\ref{eq:GSTheta2}) requires a boundary condition for $\Theta$; this is related to the question of  boundary conditions for $\vr$.

We scale the system as follows: the length $s$ is scaled with respect to $3 \nu/\vn$, and the velocity $v$ with respect to $\vn$. Then, (\ref{eq:GSTheta1}), (\ref{eq:GSTheta2}), (\ref{eq:Defxi}), (\ref{eq:bcVAtNozzle}), (\ref{eq:bcVAtBelt}) and (\ref{eq:YConstr}) become
\begin{eqnarray}
  &\ & \xi_s = \frac{A \sin(\Theta)}{v}, \label{eq:GSsTheta1} \\
  &\ & \Theta_s = \frac{ A\cos (\Theta)}{\xi v}, \label{eq:GSsTheta2} \\
  &\ & \xi=v-\frac{v_s}{v}, \label{eq:GSsTheta3} \\
  &\ & v(0)=1, \label{eq:GSsTheta4} \\
  &\ & v(\se)=\Dr, \label{eq:GSsTheta5} \\
  &\ & \int_0^{\se} \sin(\Theta(s)) ds=\B. \label{eq:GSsTheta6}
\end{eqnarray}
Here, $A=3g\nu/\vn^3$, $\B=\vn L/(3\nu)$ is the Reynolds number, $\Dr=\vb/\vn$ is the draw ratio, and the scaled $\se$  becomes $\se\vn/(3\nu)$. The dimensionless number $A$ is related to the Froude number $\mathrm{Fr}=\vn/\sqrt{gL}$ and $\B$ as $A=1/(\B\mathrm{Fr}^2)$.
After scaling the system is described in terms of three positive dimensionless numbers, which define a parameter space $\V$ as
\be
 \V= \{(A,\B,\Dr):A>0,\B>0,\Dr>0\} \label{eq:parSpace}.
\ee
The nozzle orientation $\na$ only appears in the boundary condition for $\Theta$ for the concave jet (\ref{eq:bcThetaConcave}), and is considered to be fixed.

By replacing the material coordinate $s$ by the time variable $\tm$, according to
\be
 ds= v(\tm) d\tm, \label{eq:chng_st}
\ee
the system (\ref{eq:GSsTheta1})-(\ref{eq:GSsTheta6}) becomes
\begin{eqnarray}
  &\ &    \xi_\tm = A \sin(\Theta), \label{eq:xTtau1} \\
  &\ &   \Theta_\tm = \frac{ A\cos (\Theta)}{\xi}, \label{eq:xTtau2} \\
  &\ &   \xi=v-\frac{v_\tm}{v^2}, \label{eq:xTtau3} \\
  &\ &   v(0)=1, \label{eq:xTtau4} \\
  &\ &   v(\tme)=\Dr, \label{eq:xTtau5} \\
  &\ & \int_0^{\tme} \sin(\Theta(\tm)) v(\tm) d \tm=\B. \label{eq:xTtau6}
\end{eqnarray}
Here, $\tme$ is the result of the coordinate transformation (\ref{eq:chng_st}) of $\se=\int_0^{\tme} v(\tm)d\tm$.
Next, we solve (\ref{eq:xTtau1}) and (\ref{eq:xTtau2}), using the first integral
\be
 \xi\sin(\Theta)=A\tm+c_1,
\ee
to obtain
\begin{eqnarray}
 \xi &=& \pm \sqrt{A^2\tm^2+2 A c_1 \tm+c_2}, \label{eq:solxi} \\
 \Theta &=& \pm \arcsin \left(\frac{A \tm+ c_1}{\sqrt{A^2\tm^2+2 A c_1 \tm+c_2}}\right). \label{eq:solTheta}
\end{eqnarray}
Here, $c_1$ and $c_2$ are unknown constants to be determined later.

In the analysis we restrict ourselves to solutions with $\Theta\in[0,\pi/2]$. Then, we conclude  from (\ref{eq:xTtau1}) that $\xi$ is a strictly increasing function. Therefore, we distinguish three possible situations for the sign of $\xi$: always positive, a sign change from negative to positive, and always negative, i.e.
\begin{eqnarray}
 0<\xi(0)<\xi(\tme),      \label{eq:xiconcave}\\
 \xi(0)\le0\le\xi(\tme),  \label{eq:xivertical} \\
 \xi(0)<\xi(\tme)<0.      \label{eq:xiconvex}
\end{eqnarray}

If (\ref{eq:xiconcave}) holds, then it follows from (\ref{eq:xTtau2}) that $\Theta$ is a strictly increasing function for $\Theta<\pi/2$, implying that the jet has a concave shape. As will be justified in Section~\ref{sec:ModelBC} we prescribe the nozzle orientation angle as the boundary condition for $\Theta$, i.e.
\be
\Theta(0)=\na. \label{eq:bcThetaConcave}
\ee
Substitution of (\ref{eq:bcThetaConcave}) into (\ref{eq:solxi})-(\ref{eq:solTheta}) gives
\begin{eqnarray}
        \xi &=& \sqrt{A^2\tm^2+2 A \sqrt{c_2}\sin(\na) \tm+c_2}, \label{eq:solxiIR} \\
        \Theta &=& \arcsin \left(\frac{A \tm+ \sqrt{c_2}\sin(\na)}{\sqrt{A^2\tm^2+2 A \sqrt{c_2}\sin(\na) \tm+c_2}}\right). \label{eq:solThetaIR}
\end{eqnarray}
Because (\ref{eq:xiconcave}) implies a concave shape, we refer to a jet satisfying (\ref{eq:xiconcave}) as a convex jet.

For (\ref{eq:xivertical}) to hold, there must exist a $\tm^*\in[0,\tme]$ such that $\xi(\tm^*)=0$. Then from (\ref{eq:xTtau2}), it follows that $\Theta(\tm^*)=\pi/2$. Substituting $\tm^*$ into (\ref{eq:solTheta}), we have
\be
   \frac{A \tm^*+ c_1}{\sqrt{A^2(\tm^*)^2+2 A c_1\tm^*+c_2}}=1,
\ee
giving $c_1^2=c_2$. This implies that
\be
\Theta\equiv \pi/2,
\ee
for all $\tm\in[0,\tme]$, and hence the jet is vertical, and
\be
\xi=A \tm+c_1.
\ee
For $\xi$ obeying (\ref{eq:xivertical}), we obtain $\xi(\tm)=A\tm-\sqrt{(c_2)^2}$. Because (\ref{eq:xivertical}) implies a vertical shape, we refer to a jet satisfying (\ref{eq:xivertical}) as a vertical jet. Note that for the vertical jet, as will be shown in Section~\ref{sec:ModelBC}, no boundary condition for $\Theta$ is necessary.

If (\ref{eq:xiconvex}) holds, then it follows from (\ref{eq:xTtau2}) that $\Theta$ is a strictly decreasing function for $\Theta<\pi/2$. In this case the jet has a convex shape. As will be justified in Section~\ref{sec:ModelBC}, we require tangency for the jet at the belt, i.e.
\be
    \Theta(\tme)=0. \label{eq:bcThetaConvex}
\ee
 Then
\begin{eqnarray}
        \xi &=& - \sqrt{A^2\tm(\tm-2\tme)+c_2}\, , \label{eq:solxiVR} \\
        \Theta &=& \arcsin \left(\frac{A (\tme-\tm)}{\sqrt{A^2\tm(\tm-2\tme)+c_2}}\right). \label{eq:solThetaVR}
\end{eqnarray}
Because (\ref{eq:xiconvex}) implies a convex shape we call a jet for which (\ref{eq:xiconvex}) holds a convex jet.

By substituting the found solutions for $\xi$ and $\Theta$ into (\ref{eq:xTtau3})-(\ref{eq:xTtau6}) for the three situations (\ref{eq:xiconcave})-(\ref{eq:xiconvex}) we successively obtain
\begin{eqnarray}
  v-\frac{v_\tm}{v^2}&=&\left\{
        \begin{array}{ll}
            \sqrt{A^2\tm^2+w^2+2 A\tm w\sin(\na)} & \mbox{concave jet,}\\
            w+A\tm & \mbox{vertical jet,}\\
            w\sqrt{A^2\tm(\tm-2\tme)/w^2+1}& \mbox{convex jet,}\\
        \end{array}
       \right. \label{eq:eqv1} \\
  v(0)&=&1, \label{eq:eqv2} \\
  v(\tme)&=&\Dr,  \label{eq:eqv3} \\
  \B&=&\left\{
        \begin{array}{ll}
            \int_0^{\tme} \frac{A \tm+ w\sin(\na)}{\sqrt{A^2\tm^2+w^2+2 A\tm w\sin(\na)}} v(\tm) d \tm & \mbox{concave jet,}\\
            \int_0^{\tme}  v(\tm) d \tm & \mbox{vertical jet,}\\
            \int_0^{\tme} \frac{A (\tme-\tm)}{\sqrt{A^2\tm(\tm-2\tme)+w^2}} v(\tm) d \tm & \mbox{convex jet,}\\
        \end{array}
       \right. \label{eq:eqv4}
\end{eqnarray}
where $w=\xi(0)$. We refer to the situations of concave, vertical and convex jets as concave, vertical and convex flow regimes, respectively.

For given $w\in\mathbb{R}$ and flow regime, the problem (\ref{eq:eqv1})-(\ref{eq:eqv3}) has a solution $\vtw$ and $\tmew$, where $\tmew$ satisfies (\ref{eq:eqv3}). Here, we assume that for any $w$, (\ref{eq:eqv3}) has only one solution, which is not always true. However, this allows us to illustrate a solution procedure.

Substituting $\vtw$ and $\tmew$ into the integrals (\ref{eq:eqv4}), we obtain the functions of~$w$:
\be
        \begin{array}{ll}
            \displaystyle\Ic(w)=\int_0^{\tmew} \frac{A \tm+ w\sin(\na)}{\sqrt{A^2\tm^2+w^2+2 A\tm w\sin(\na)}} \vctw d \tm & \mbox{concave jet,}\\
            \displaystyle\Iv(w)=\int_0^{\tmew}  \vvtw d \tm & \mbox{vertical jet,}\\
           \displaystyle \Ix(w)=\int_0^{\tmew} \frac{A (\tmew-\tm)}{\sqrt{A^2\tm(\tm-2\tmew)+w^2}} \vxtw d \tm & \mbox{convex jet.}\\
        \end{array}
       \label{eq:defofIw}
\ee
Here, we denote by $\vctw$, $\vvtw$ and $\vxtw$ the solution of (\ref{eq:eqv1}) for a concave, vertical and convex jet, respectively. According to (\ref{eq:xiconcave})-(\ref{eq:xiconvex}), $\Ix(w)$ and $\Iv(w)$ are defined for $w\le0$, and $\Ic(w)$ for $w>0$.
With (\ref{eq:defofIw}), solving (\ref{eq:eqv1})-(\ref{eq:eqv4}) is equivalent to solving the algebraic equation
\be
    \I_?(w)=\B,       \label{eq:eqnIweqB}
\ee
where $?$ stands for an unknown jet flow regime. Therefore, a study of existence and uniqueness of a jet solution results into a study of the existence and uniqueness of a solution to the algebraic equation (\ref{eq:eqnIweqB}).

\begin{figure}
\centering
 \includegraphics[width=0.8\textwidth]{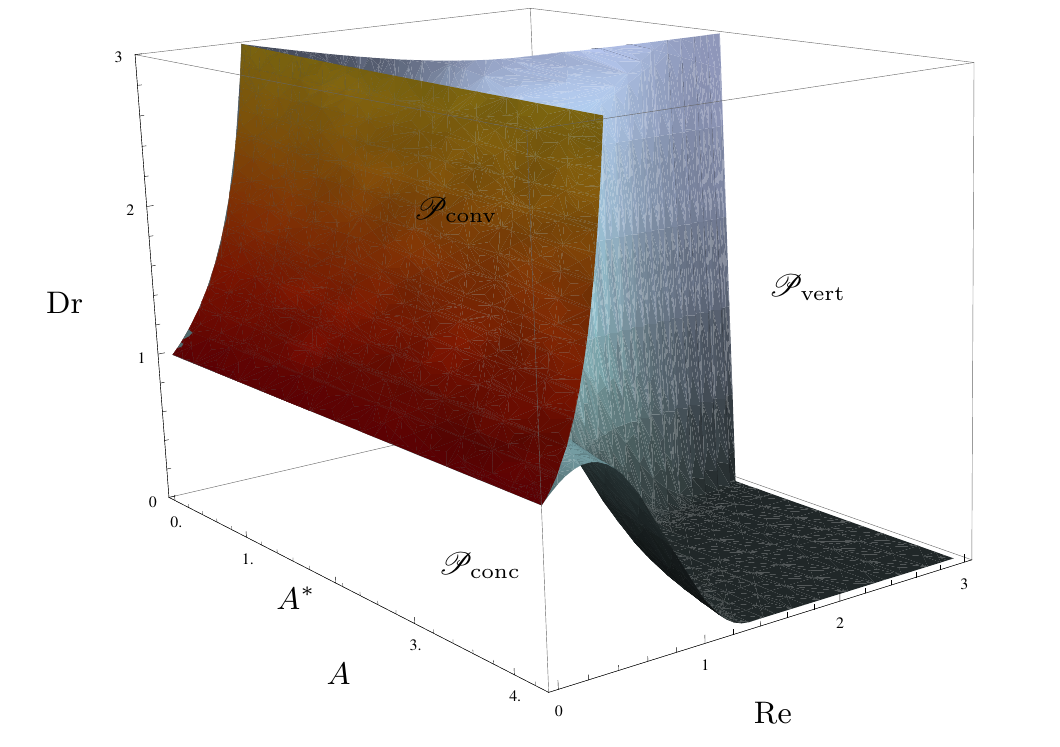}
\caption{Parameter regions for three flow regimes $\Vc$, $\Vv$ and $\Vx$. The  border between $\Vc$ and $\Vv$ is $S_1$ and the  border between $\Vx$ and $\Vv$ is $S_2$.}
   \label{fig:ThreeRegion}
\end{figure}

At this point, we like to briefly recapitulate the main steps in our solution procedure. We do this, as an example for the concave flow; the other cases are completely analogous.
The steps are:
\begin{enumerate}
 \item Solve $v=\vctw$ from $\mbox{(\ref{eq:eqv1})}_\mathrm{1}$, with use of the boundary condition (\ref{eq:eqv2}).
 \item Find $\tmew$ from (\ref{eq:eqv3}) as $v_\mathrm{conc}(\tmew;w)=\Dr$.
 \item Calculate $\Ic(w)$ from (\ref{eq:defofIw}).
 \item Solve $w$ from (\ref{eq:eqnIweqB}).
\end{enumerate}
The partitioning of the parameter space $\V$ into the regions of concave $\Vc$, vertical $\Vv$ and convex $\Vx$ jets is presented in Figure~\ref{fig:ThreeRegion}.
The partitioning follows from the solutions of (\ref{eq:eqv1})-(\ref{eq:eqv4}) with the additional condition $\xi(\tme)=0$ for the border between $\Vv$ and  $\Vx$, and $\xi(0)=0$ for the border between $\Vc$ and  $\Vv$.

For any set of parameters from $\Vc$, or $\Vv$, or $\Vx$, a corresponding solution of (\ref{eq:eqv1})-(\ref{eq:eqv4}) exists. This solution is unique in the case of vertical and convex flow. For the concave flow a solution might not be unique when the nozzle does not point vertically down and $\Dr>1$. In this case up to two solutions exist for a concave jet and one solution for a convex or vertical jet. The questions of existence and uniqueness are fully treated in the upcoming paper \cite{Hlod4}, and for the convex jet in \cite{Hlod1}.

In the next section we justify our choice of the boundary conditions for $\Theta$, i.e. (\ref{eq:bcThetaConcave}) and (\ref{eq:bcThetaConvex}), for concave and convex flow, and we explain that no boundary condition is needed for $\Theta$ in case of vertical flow.
\section{Justification of boundary conditions for $\Theta$} \label{sec:ModelBC}
In this section we explain our choice of boundary conditions for $\Theta$ made in the previous section. We use the highest-order part, or principal part, of the conservation of momentum equation (\ref{eq:consMom})
\be
 \vr_{tt}+2 v \vr_{st}+v \xi \vr_{ss}=[\vr_{tt}++2 v \vr_{st}+v^2 \vr_{ss}]-v_s \vr_{ss}  \label{eq:CMHx}
\ee
This equation is of hyperbolic type in the neighborhood of $s=\se$ if $\xi(\se)$ is close to zero, and in the neighborhood of $s=0$, if $\xi(0)$ is close to zero, because $v_s=v2+\xi v>0$ in theses situations. We consider the direction of the characteristics of (\ref{eq:CMHx}), either at $s=\se$ if  $\xi(\se)$ changes sign, or at $s=0$ if  $\xi(0)$ changes sign. This directly yields the number of boundary conditions, which must be prescribed at $s=\se$ or $s=0$. The reason is that the number of boundary conditions at any point of the boundary is equal to the number of characteristics pointing inside the domain at this point; \cite[see][p 417]{NumAprHypSys}.

The characteristics equation \cite[see][p 57]{Davis} for (\ref{eq:CMHx}) is
\be
z^2-2 v z+ v^2-v_s=0, \label{eq:chareq}
\ee
where $z$ represents the velocity of a characteristic. Equation (\ref{eq:chareq}) has the solutions
\be
 z_1= v+\sqrt{v_s}, \ \ \
   z_2= v-\sqrt{v_s}. \label{eq:chareqsol}
\ee
According (\ref{eq:chareqsol}) and (\ref{eq:GSsTheta3}), the directions of the characteristics of (\ref{eq:CMHx}) depend on the sign of $\xi$ as follows:
\begin{enumerate}
\item  If $\xi>0$ then $z_1>0$ and $z_2>0$, i.e. the two characteristics point to the right.
\item  If $\xi=0$ then $z_1>0$ and $z_2=0$, i.e. one characteristic points to the right and one is stationary.
\item  If $\xi<0$ then $z_1>0$ and $z_2<0$, i.e. one characteristic points to the left and one to the right.
\end{enumerate}
From the characterization of the flow regimes (\ref{eq:xiconcave})-(\ref{eq:xiconvex}) we infer that:
\begin{itemize}
\item At $s=0$ two boundary conditions for $\vr(s,t)$ are necessary in case of  a concave jet ($\xi(0)>0$), and only one in case of  a vertical or convex jet ($\xi(0)\le0$).
\item
In case of a convex jet, one boundary condition for $\vr$ is necessary at $s=\se$ ($\xi(\se)<0$), and no in case of vertical or concave jets ($\xi(\se)\ge0$).
\end{itemize}
For all three situations we prescribe the nozzle position. In addition, for the concave jet the nozzle orientation is prescribed by (\ref{eq:bcThetaConcave}), and for the convex jet the tangency condition (\ref{eq:bcThetaConvex}) with the belt is prescribed. This justifies our choice of boundary conditions  (\ref{eq:bcThetaConcave}) and (\ref{eq:bcThetaConvex}) for the stationary problem.

The analysis of characteristics, as directions of information propagation, explains why the nozzle orientation influences the jet shape only in the case of  concave flow, and why the belt orientation influences the jet shape only in case of convex flow.
\begin{itemize}
\item In concave flow all information about the jet shape travels from the nozzle to the belt. Therefore, not only nozzle position but also nozzle orientation is relevant for the jet. In addition, no information on angle travels back from the belt.
\item In vertical flow only one characteristic (at the nozzle) points inside the domain. Therefore, no information about nozzle orientation or belt movement direction influences the jet shape. Thus, in vertical flow the nozzle and the belt orientations  are irrelevant for the jet.
\item In convex flow one characteristic points inside the domain at the nozzle and one at the belt. Hence, information about the direction of the belt movement influences the jet shape, and therefore, the belt orientation becomes relevant in convex flow.
\end{itemize}
\section{Results from the model} \label{sec:ResultsFromModel}
In this section we present some results from our model. We analyse the partitioning of the parameter space. Next, we investigate changes of the flow type if one of the physical parameters ($L$, $\nu$, $\vb$, $\vn$) is varied. We describe the trajectories of the process parameters in the parameter space $\V$, and we illustrate the jet shape evolution. Note that the only possible transitions between flow types are between $\Vv$  and  $\Vx$, and between $\Vv$  and  $\Vc$; see Figure~\ref{fig:ThreeRegion}.

\begin{figure}
   \center
   \includegraphics[width=0.9\textwidth]{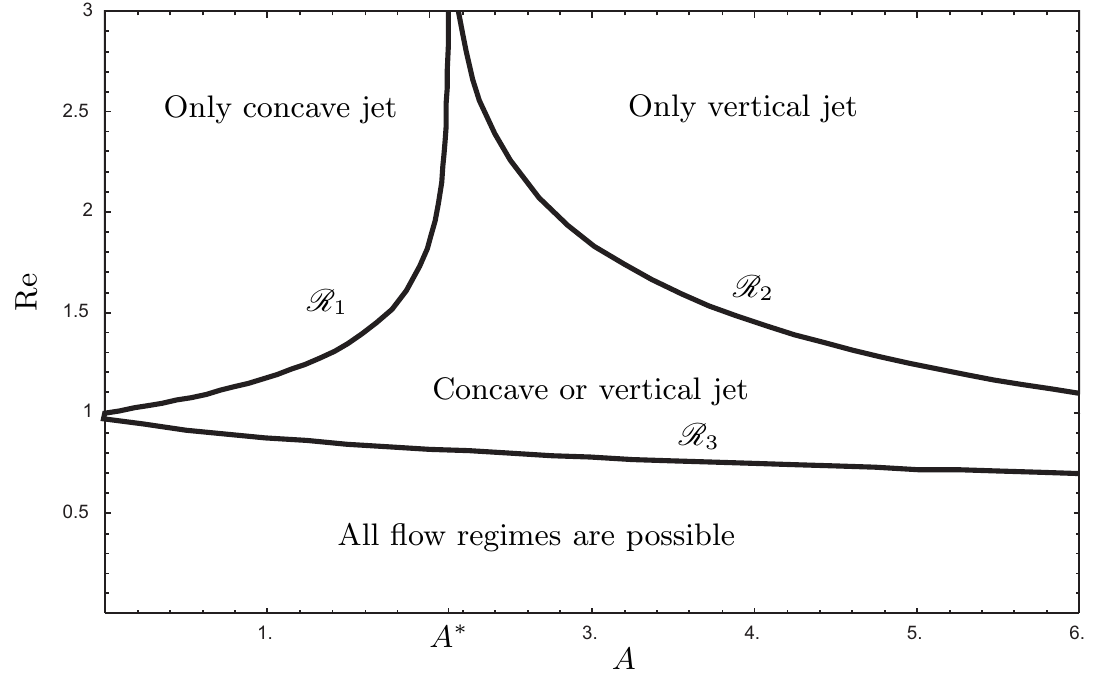}
   \caption{Regions of $A$ and $\B$ for which, for all $\Dr$, different flow regime exist. In the region $\{ A<\Acr,\, \B>\msR_1(A)\}$, only concave jets are possible, in $\{ A>\Acr,\, \B>\msR_2(A)\}$ only vertical jets, while in  $\{ A>0,\, \B<\msR_3(A)\}$ all flow regimes are possible. In the remaining region between $\msR_1$, $\msR_2$ and $\msR_3$ concave and vertical jets, but no convex jets, are possible.}
   \label{fig:OnlyIorVI}
\end{figure}
The projection of the regions for the three flow regimes onto the $(A,\B)$-plane is depicted in Figure~\ref{fig:OnlyIorVI}, and following is valid for all values of $\Dr$. We observe a region  $\{ A<\Acr,\, \B>\msR_1(A)\}$ where the jet is concave, and a region $\{ A>\Acr,\, \B>\msR_2(A)\}$ where it is vertical. In the region between $\msR_1$, $\msR_2$ and $\msR_3$ concave or vertical flow  is possible, but there can be no convex flow. Finally, in the region $\{ A>0,\, \B<\msR_3(A)\}$, all three flow regimes are possible. Hence, it is only in the latter region, where $\B<\msR_3(A)<1$, that a convex jet can occur.

\begin{figure}
   \center
   \includegraphics[width=0.9\textwidth]{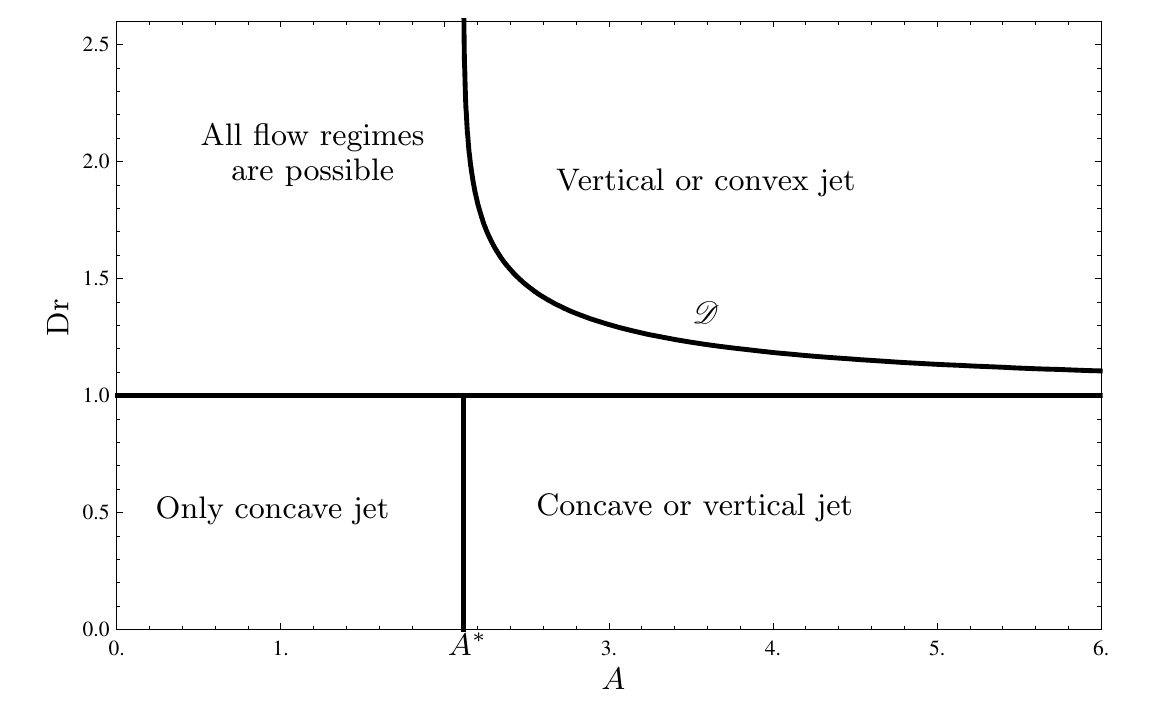}
   \caption{Regions of $A$ and $\Dr$ for which for all $\B$ different flow regimes exist. In the region $\{A<\Acr,\, \Dr<1\}$ only concave jets are possible, in $\{A>\Acr,\, \Dr<1\}$ concave and vertical jets (no convex jets), and in $\{A>\Acr,\, \Dr>\msD(A)\}$ convex and vertical jets (no concave jets). In the remaining region between $\Dr=1$ and $\msD$, all flow regimes are possible.}
   \label{fig:OnlyVbAPl}
\end{figure}
The parameter regions projection onto the  $(A,\Dr)$-plane is depicted in Figure~\ref{fig:OnlyVbAPl}; and following holds for all $\B$. We observe that for $\{A>\Acr,\, \Dr<1\}$ only a concave jet is possible, while concave or vertical flow is possible for $\{A>\Acr,\, \Dr<1\}$. In the region $\{A>\Acr,\, \Dr>\msD(A)\}$ convex or vertical flow is possible, while in the rest of region $\{\Dr>1\}$ all three flow regimes are possible. Hence, a convex jet can only occur if $\Dr>1$.
\begin{figure}
\centering
\includegraphics[width=0.7\textwidth]{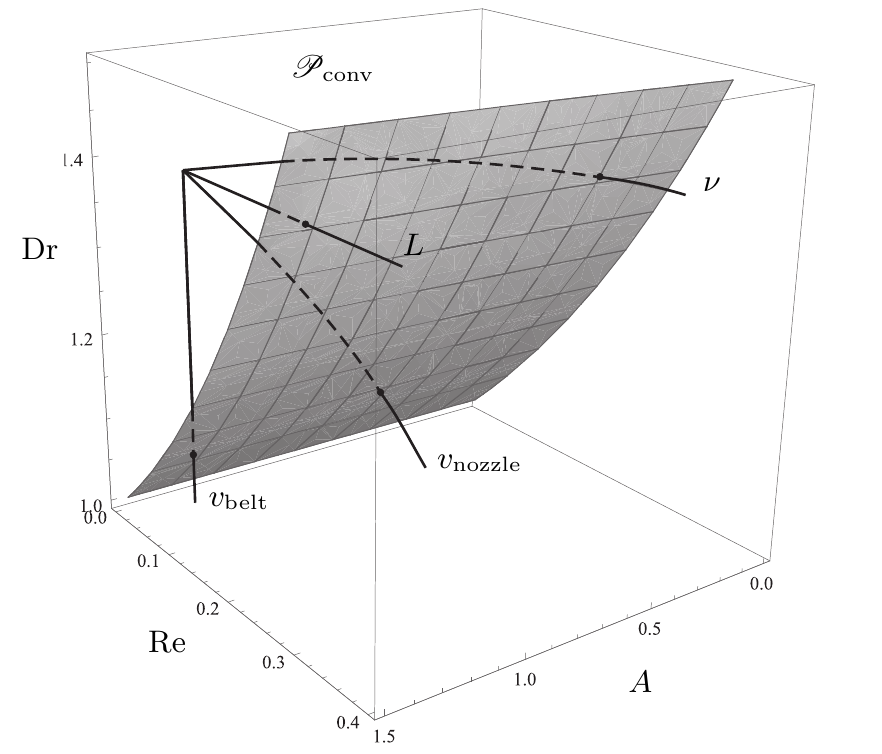}
\caption{Traces of the point $(A,\B,\Dr)$ as we change one of the physical parameters ($L$ increases, $\nu$ decreases, $\vb$ decreases, $\vn$ increases). The curves originate at the same point in the region $\Vx$ and eventually leave $\Vx$ by crossing the separating gray surface at the points indicated by the dots.}
 \label{fig:VJetRegion1}
\end{figure}
\begin{figure}
\centering
\subfigure[Convex jet shapes for different $L$: 1, 1.7, 2.2, and 3.5 $\mathrm{cm}$. The shape approaches the vertical as $L$ increases.  ]{\label{subfig:ChLVJ}\includegraphics[width=0.47\textwidth]{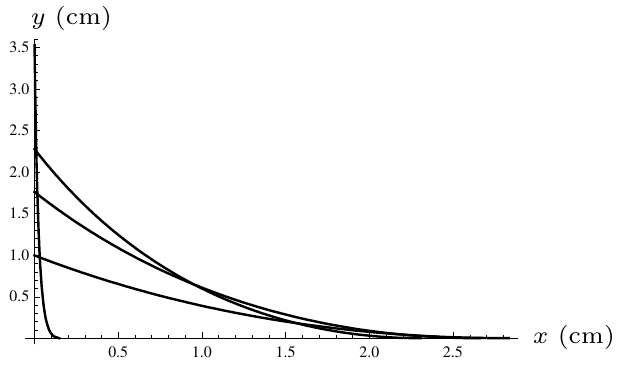}}\ \
\subfigure[Convex jet shapes for different $\nu$: 0.047, 0.026, 0.015, and 0.012 $\mathrm{m^2/s}$. The shape approaches the vertical as $\nu$ decreases.  ]{\label{subfig:ChmuVJ}\includegraphics[width=0.47\textwidth]{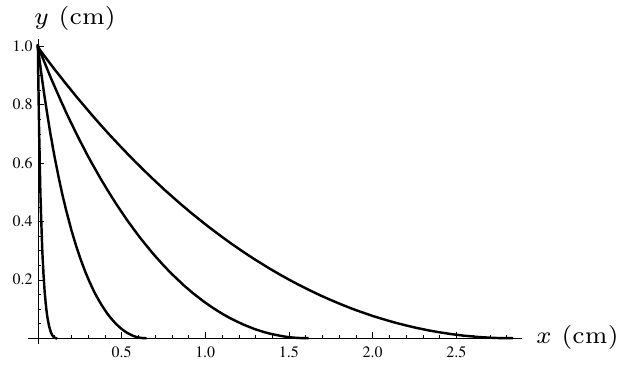}}\\
\subfigure[Convex jet shapes for different $\vb$: 1.4, 1.21, 1.11, and 1.08 $\mathrm{m/s}$. The  shape approaches the vertical as $\vb$ decreases.  ]{\label{subfig:ChvbVJ}\includegraphics[width=0.47\textwidth]{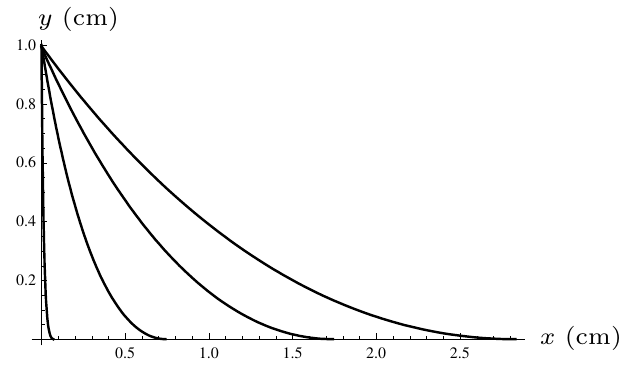}}\ \
\subfigure[Convex jet shapes for different $\vn$: 1, 1.16, 1.24, and 1.26 $\mathrm{m/s}$. The shape approaches the vertical as $\vn$ increases.  ]{\label{subfig:ChvnVJ}\includegraphics[width=0.47\textwidth]{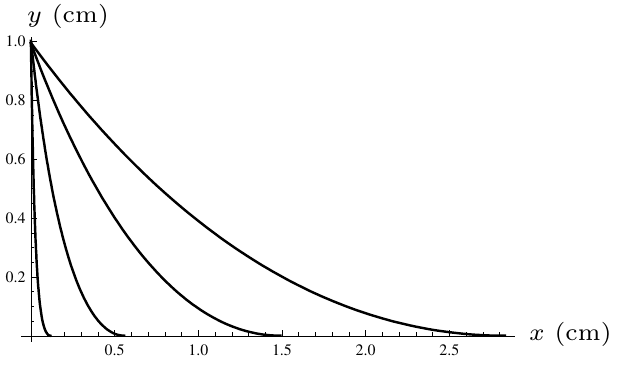}}
\caption{Shapes of the convex jet for different values of $L$, $\nu$, $\vb$ and $\vn$. The reference values are $L=1\,\mathrm{cm}$, $\nu=0.047\,\mathrm{m^2/s}$, $\vb=1.4\,\mathrm{m/s}$, and $\vn=1\,\mathrm{m/s}$.}
  \label{fig:ChVJ}
\end{figure}

Next, we study the evolution of the jet if one of the physical parameters varies as to change the flow type from convex to vertical. For a reference configuration we take the physical parameters $L=1\,\mathrm{cm}$, $\nu=0.047\,\mathrm{m^2/s}$, $\vb=1.4\,\mathrm{m/s}$, and $\vn=1\,\mathrm{m/s}$, for which the jet is convex. Then, if we increase $L$,  decrease $\nu$, decrease $\vb$, or increase $\vn$, eventually the jet flow changes from convex to vertical. The corresponding curves in the parameter space $\V$ are indicated in Figure~\ref{fig:VJetRegion1}.

Changes of the jet shape while only one of the physical parameters $L$, $\nu$, $\vb$, or $\vn$ varies as described above are shown in Figures \ref{subfig:ChLVJ}, \ref{subfig:ChmuVJ}, \ref{subfig:ChvbVJ}, and \ref{subfig:ChvnVJ}, respectively. In Figures~\ref{fig:VJetRegion1}~and~\ref{fig:ChVJ} we see that if the point $(A,\B,\Dr)$ approaches the boundary of $\Vx$, the jet shape becomes vertical. If  $(A,\B,\Dr)$ is very close to the boundary of $\Vx$ the jet shape is almost vertical, except for the small region near the belt where the jet rapidly bends to the horizontal belt direction.

\begin{figure}
\centering
\includegraphics[width=0.7\textwidth]{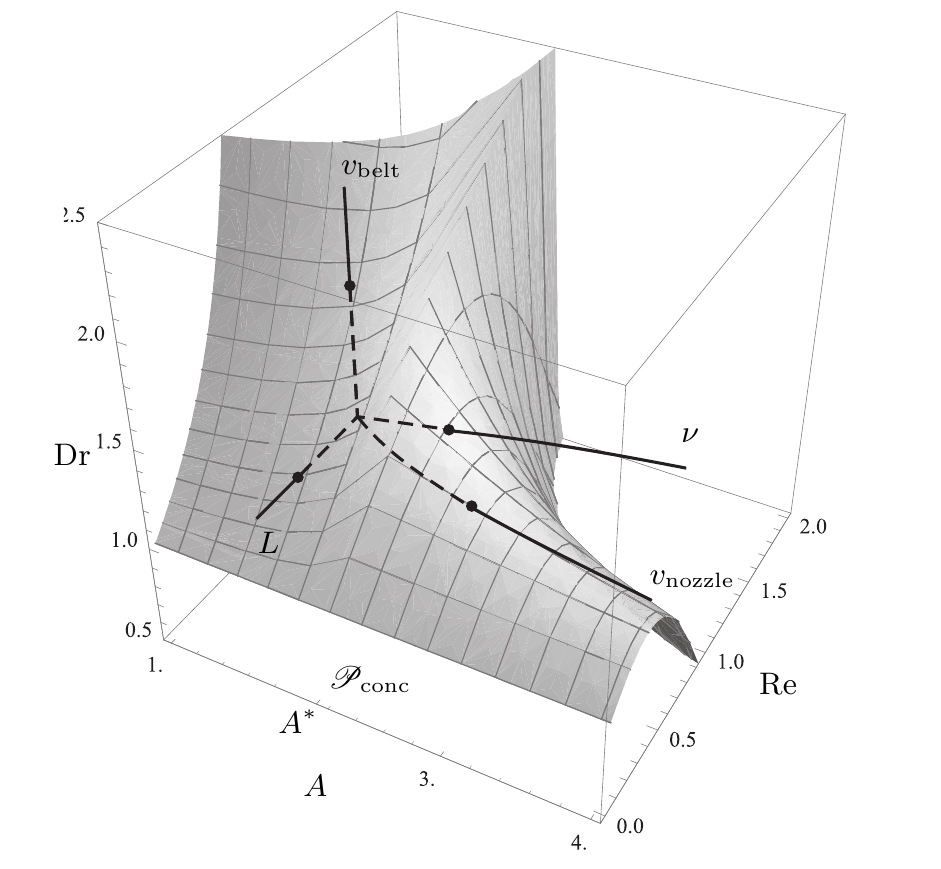}
\caption{Curves in the parameter space $\V$ as we change one of the parameters ($L$ decreases, $\nu$ increases, $\vb$ increases, $\vn$ decreases). The curves originates at the same point in the region $\Vc$ and eventually leave $\Vc$ by crossing the separating surface at the points indicated by the dots.}
  \label{fig:IJetRegion1}
\end{figure}
\begin{figure}
\centering
\subfigure[Concave jet shapes for different $L$: 30, 18, 13, and 12 $\mathrm{cm}$. The shape approaches the vertical as $L$ decreases.]{\label{subfig:ChLIJ}\includegraphics[width=0.45\textwidth]{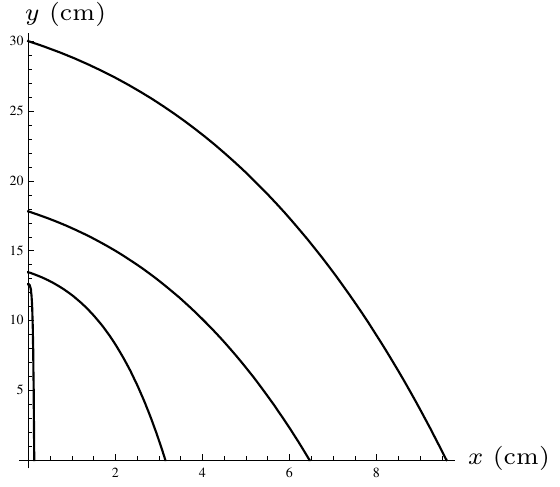}}\ \
\subfigure[Concave jet shapes for different $\nu$: 0.2, 0.26, 0.3, and 0.32 $\mathrm{m^2/s}$. The shape approaches the vertical as $\nu$ increases.]{\label{subfig:ChmuIJ}\includegraphics[width=0.45\textwidth]{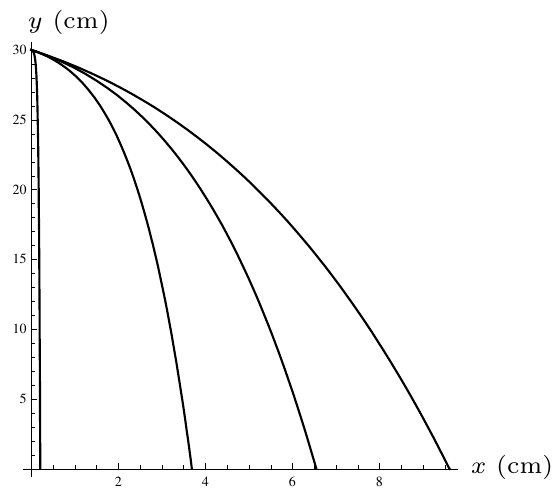}}\\
\subfigure[Concave jet shapes for different $\vb$: 2, 2.57, 2.86, and 2.95 $\mathrm{m/s}$. The shape approaches the vertical as $\vb$ increases]{\label{subfig:ChvbIJ}\includegraphics[width=0.45\textwidth]{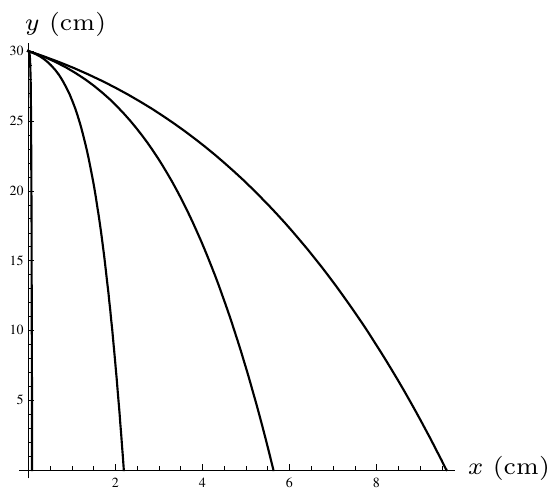}}\ \
\subfigure[Concave jet shapes for different $\vn$: 1.5, 1.4, 1.36, and 1.34 $\mathrm{m/s}$. The shape approaches the vertical as $\vn$ decreases.]{\label{subfig:ChvnIJ}\includegraphics[width=0.45\textwidth]{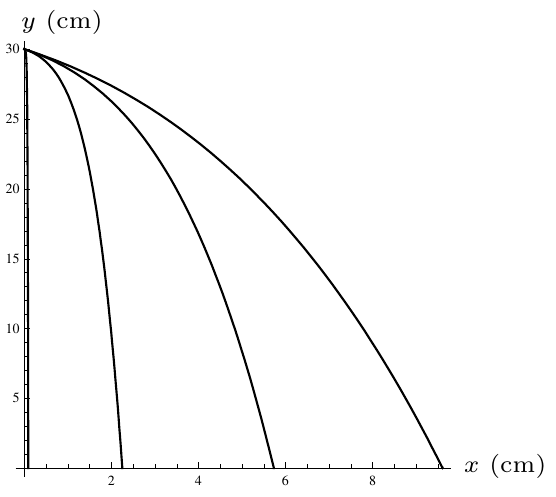}}
\caption{Shapes of the concave jet for different values of $L$, $\nu$, $\vb$, $\vn$. The reference values are $L=30\,\mathrm{cm}$, $\nu=0.2\,\mathrm{m^2/s}$, $\vb=2\,\mathrm{m/s}$, and $\vn=1.5\,\mathrm{m/s}$. The nozzle orientation is $\na=\pi/4$.}
  \label{fig:ChIJ}
\end{figure}
The analysis of the parameter region for the concave jet is more complex than that for the convex jet.  In case $\{A>\Acr,\ \Dr<1\}$ (``Only concave jet'' in Figure~\ref{fig:OnlyVbAPl}), the flow is concave for all $L$. Similar, if $\{A<\Acr,\ \B>\mathscr{R}_1(A)\}$ (``Only concave jet'' in Figure~\ref{fig:OnlyIorVI}), the flow is concave for all $\vb$. In a situation when $\nu$ decreases or $\vn$ increases, $A$ approaches zero and $\B$ approaches infinity since $A=3g\nu/\vn^3$, and $\B=\vn L/(3\nu)$. Thus, eventually the point $(A,\B)$ enters the  ``Only concave jet'' region in Figure~\ref{fig:OnlyIorVI}. Hence, if the jet is not in the concave flow regime, decreasing $\nu$ or increasing $\vn$  makes the jet to become concave eventually.

To illustrate the change of flow from concave to vertical, while only one of the parameters $L$, $\nu$, $\vb$, and $\vn$ varies, we take the reference values $L=30\,\mathrm{cm}$, $\nu=0.2\,\mathrm{m^2/s}$, $\vb=2\,\mathrm{m/s}$, and $\vn=1.5\,\mathrm{m/s}$. Then, if we decrease $L$, increase $\nu$, increase $\vb$, or decrease $\vn$ eventually the jet flow changes from concave to vertical. The curves in the parameter space $\V$ are indicated in Figure~\ref{fig:IJetRegion1}.

Changes of the jet shape for  $\na=\pi/4$, while only one of the physical parameters  $L$, $\nu$, $\vb$, or $\vn$ varies as described above are shown in Figures~\ref{subfig:ChLIJ}, \ref{subfig:ChmuIJ}, \ref{subfig:ChvbIJ}, and \ref{subfig:ChvnIJ}, respectively. In Figures~\ref{fig:IJetRegion1}~and~\ref{fig:ChIJ} we see that if the point $(A,\B,\Dr)$ approaches the boundary of $\Vc$, the jet shape becomes more vertical. If  $(A,\B,\Dr)$ is very close to the boundary of $\Vc$ the jet shape is almost vertical except for the small region near the belt where the jet rapidly bends from the nozzle direction to an almost vertical one.

The analysis above shows that the transition between the convex and the concave flow regimes as parameters continuously vary is only possible via the vertical flow. In the next section we compare the results from our model with those from our experiments.
\section{Comparison between the model and experiments}  \label{sec:Comparison}

In this section we validate our model using the results of the experiments described in Section \ref{sec:experiments} by comparing the corresponding relations between $\xe$ and $\vb$. We compare the shapes from the experiments and the model, and discuss differences and similarities for jets in convex and concave flow regimes.
\subsection{Comparison of $\xe$}
\begin{figure}
\centering
\includegraphics[width=0.6\textwidth]{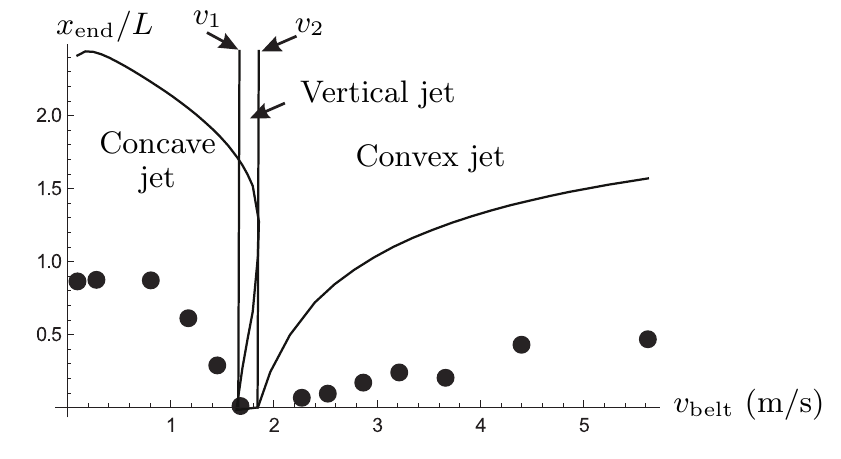}
\caption{Comparison of the relations between $\vb$ and $\xe$ as obtained from the model $\Cxe$ (solid line) and from the experiment (dots) for
 $L=4.1\,\mathrm{cm}$, $\nu=0.047\,\mathrm{m^2/s}$, $\vn=1.147\,\mathrm{m/s}$, $\D=0.4\,\mathrm{mm}$, and $\na=5^\circ$. A vertical jet occurs for $\vb\in[v_1,v_2]$, and $\xe=0$.}
  \label{fig:XendComp}
\end{figure}
We start by comparing the model predictions of the touchdown point with the ones obtained from the experiments in Figure~\ref{fig:xendpos}. The following values of parameters are used $\vn=1.147\,\mathrm{m/s}$, $\L=4.1\,\mathrm{cm}$, $\nu=0.047\,\mathrm{m^2/s}$, $\D=0.4\,\mathrm{mm}$, and $\na=5^\circ$. In Figure~\ref{fig:XendComp} we present the relations between $\xe$ and $\vb$ obtained from the model (solid curve) and the experiments (dots). The partitioning of the parameter space $\V$ gives the regions of $\vb$ for the three flow regimes as indicated by the vertical lines (at $\vb=v_1\mbox{ and }v_2$). For small $\vb$ the flow is concave, for increasing $\vb$ vertical, and for even larger $\vb$ convex.

A curve $\Cxe$, describing the relation between $\xe$ and $\vb$ as obtained from the model  (solid curve in Figure~\ref{fig:XendComp}) consists of three parts:
\begin{itemize}
\item[\textbf{Concave jet part.}]  For $\vb$ close to 0, $\xe$ increases as $\vb$ increases, till $\Cxe$ reaches its maximum. For further increasing $\vb$, $\xe$ decreases till $\Cxe$ becomes vertical (at $\vb=v_2$). After this point $\Cxe$ bends back and $\xe$ decreases to zero; during this phase $\vb$ decreases to $v_1$. At this point where $\xe$ becomes zero, the jet becomes vertical; in the preceding part of $\Cxe$ the jet is concave. Hence, for $\vb\in[0,v_2]$ and $\xe>0$, the jet is concave.
\item[\textbf{Vertical jet part.}] This part of $\Cxe$ is horizontal with $\xe=0$, while $\vb$ increases from $v_1$ to $v_2$. In this part, the flow is vertical.
\item[\textbf{Convex jet part.}] This part of $\Cxe$ starts at $\xe=0$ and $\vb=v_2$, after which both  $\xe$ and $\vb$ increase. In this part the jet is convex.
\end{itemize}
Looking at the shape $\Cxe$ in Figure~\ref{fig:XendComp} within the region $v_1<\vb<v_2$, we notice that the solution is non-unique there: two concave and one vertical solution exist there. This illustrates the non-uniqueness of the jet solution for $\na<\pi/2$.

The experimental results have qualitatively the same tendency as the theoretical ones from our model. For $\vb$ close to zero (the first three experimental values) the small maximum, or a kind of plateau, is found too and after that the values monotonically decrease till $\xe$ becomes zero. This represents the transition from the concave to the vertical jet regime. The following observation points all lie in the convex jet regime and they show a monotonic increase of $\xe$ with $\vb$. Hence, the behavior of the experimental data agrees, in general, with that predicted by the model. The only difference is (non-)monotonicity in the small region between $v_1$ and $v_2$.

Although the theoretical and experimental results agree in a qualitative sense, quantitatively significant differences are found. The values of $\xe$ predicted by the model for convex and concave flows are larger than the values obtained experimentally. We comment on this in the next section

\begin{figure}
\centering
\includegraphics[width=0.9\textwidth]{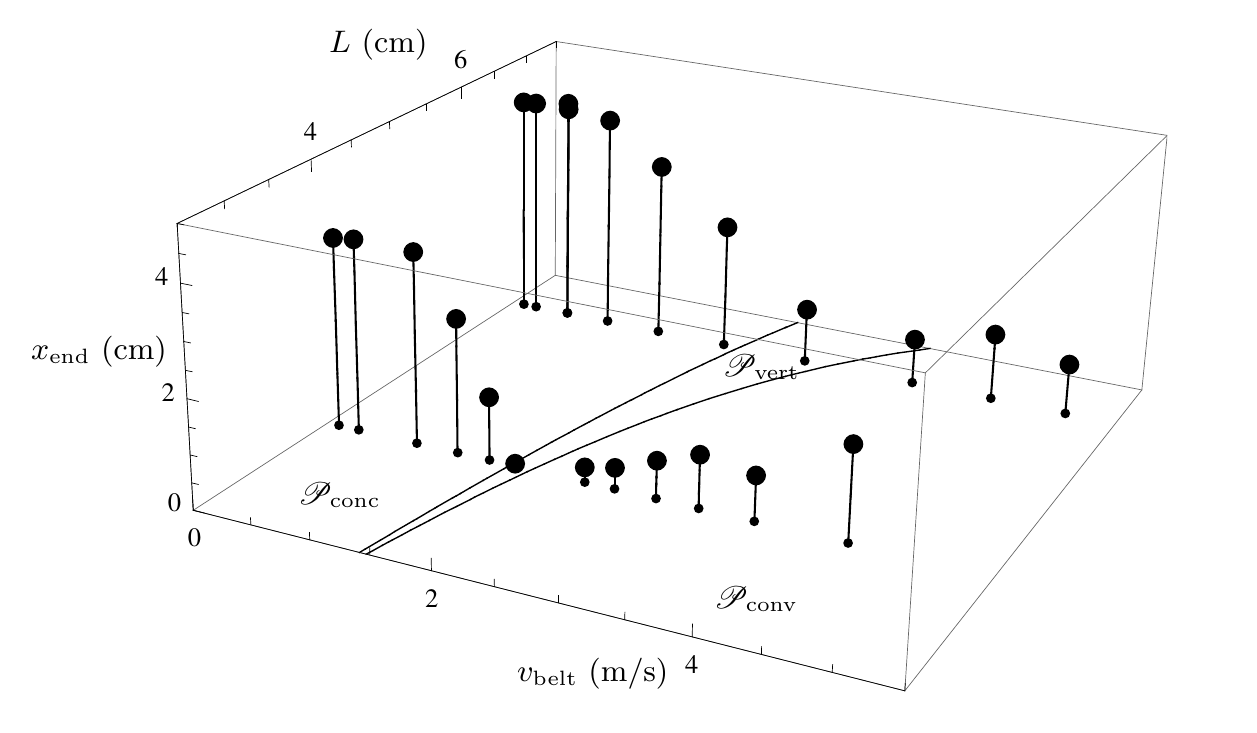}
\caption{The regions of $\vb$ and $L$ for the three flow regimes and the experimental values of $\xe$. Here, two values of $L$ are used $4.1\,\mathrm{cm}$ and $6.8\,\mathrm{cm}$.  The other parameters are $\nu=0.047\,\mathrm{m^2/s}$, $\vn=1.147\,\mathrm{m/s}$, $\D=0.4\,\mathrm{mm}$, and $\na=5^\circ$. The value of $\xe$ is small for $\vb$ in the parameter region for the vertical jet and increases if $\vb$ goes away from this region.}
  \label{fig:ResXe}
\end{figure}
Finally, we compare the predictions of the parameter regions in the $(\vb, L)$-plane for flow regimes obtained from the model with those from the experiments. To determine the type of flow regime from the experimental data, we observe the behavior of $\xe$ as $\vb$ increases: for concave flow $\xe$ decreases, for vertical flow $\xe$ is close to zero, and for convex flow $\xe$ increases. The results for two values of $L$ are presented in Figure~\ref{fig:ResXe}. For the smaller $L$ we clearly see that in the concave jet region $\xe$ decreases, stays close to zero in the vertical jet region, and increases in the convex jet region as $\vb$ increases. A similar behavior of $\xe$ is observed for the larger $L$, except in the convex jet region where the increase of $\xe$ is less significant.  Summarizing, we conclude that the experimental results for the parameter regions for the three flow regimes in the $(\vb, L)$-plane agree with the theoretical ones.
\subsection{Comparison of jet shapes}
\begin{figure}
\centering
\includegraphics[width=0.6\textwidth]{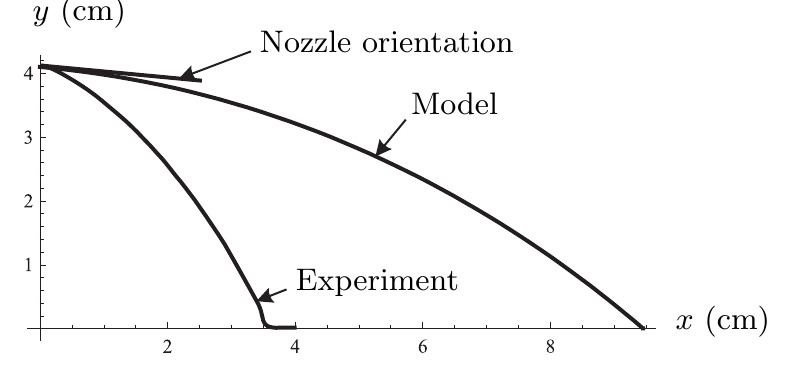}
\caption{Comparison of the concave the jet shapes obtained theoretically and experimentally for $L=4.1\,\mathrm{cm}$, $\nu=0.047\,\mathrm{m^2/s}$, $\vn=1.147\,\mathrm{m/s}$, $\vb=0.81\,\mathrm{m/s}$, and $\na=5^\circ$.}
  \label{fig:ConcJShComp}
\end{figure}
\begin{figure}
\centering
\includegraphics[width=0.6\textwidth]{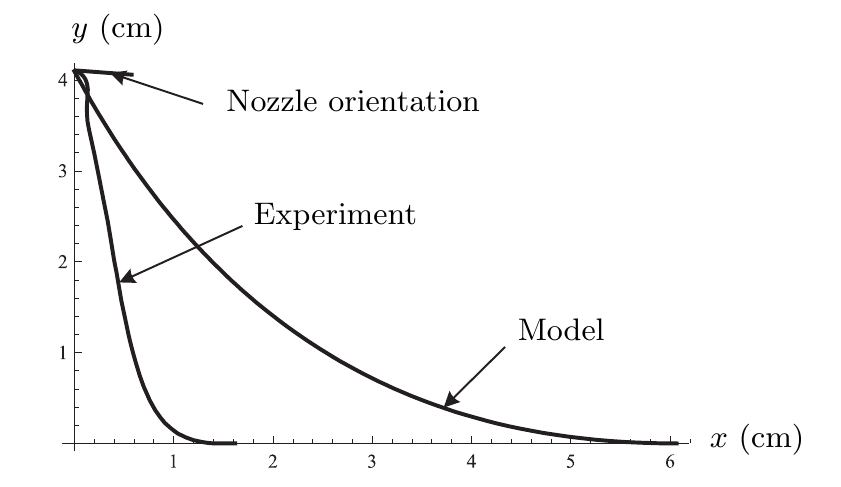}
\caption{Comparison of the convex jet shapes obtained theoretically and experimentally for $L=4.1\,\mathrm{cm}$, $\nu=0.047\,\mathrm{m^2/s}$, $\vn=1.147\,\mathrm{m/s}$, $\vb=4.398\,\mathrm{m/s}$, and $\na=5^\circ$.}
  \label{fig:ConvJShComp}
\end{figure}
Theoretically and experimentally obtained jet shapes are presented in Figures~\ref{fig:ConcJShComp}~and~\ref{fig:ConvJShComp} for a concave and a convex jet, respectively. The experimental shapes are obtained from the photos using an image analysis program.  For vertical jet the only differences between the experimental and theoretical shapes is due to bending regions near the nozzle and the belt in the experiments. Therefore, we do not discuss vertical jet here.

In Figure~\ref{fig:ConcJShComp} we compare the jet shapes for the concave flow . The experimental shape bends steeper down at the nozzle than the one from the model, and is more curved. This results in a smaller $\xe$ from the experiment than predicted theoretically. The bending region near the belt in the experimental jet does not contribute significantly  to the difference in $\xe$.

The comparison of the jet shapes for the convex flow is presented in Figure~\ref{fig:ConvJShComp}. In the experimental shape a small bending region appears near the nozzle where the jet bends from the nozzle orientation to some preferred orientation. This orientation is closer to the vertical than our theory predicts. The middle part of the experimental jet is almost straight, clearly less curved than the one obtained form the model. Near the belt the experimental jet bends to the horizontal more rapidly then the one from the model. The theoretical shape is curved more uniformly than the experimental one. All this results in a larger $\xe$ predicted theoretically than observed experimentally.
\subsection{Discussion about differences and similarities}
From comparing the jet shapes, including $\xe$, for the two flow regimes concave and convex, we observe significant differences in the jet shapes obtained theoretically and experimentally. The common difference for these flow regimes is the difference in curvature of the shapes. Also differences due to bending near the nozzle and the belt are found. The steep bending downwards near the nozzle of the concave jet in the experiment, and the differences in the jet orientation near the nozzle in the convex flow play a significant role in the observed shape differences. As a consequence the theoretical predictions of $\xe$ are larger than the experimental ones.

We conjecture that the cause of these differences lies in the effects which we did not include in our model, such as air resistance, bending stiffness and surface tension. Experiments for $\D=1\,\mathrm{mm}$, and $L=0.054\, \mathrm{m}$ are shown in Figure~\ref{fig:Experiments}. We found that the differences mentioned above in $\xe$ are smaller for the thicker jet falling from  a smaller height $L$, which makes us believe that air resistance is important. For the thicker jet, we do not observe in the experiments a steep bending of the concave jet near the nozzle; see Figures~\ref{fig:Experiment1},~\ref{fig:Experiment2}~and~\ref{fig:ExpNozzleAngle}. The effect of the bending at the nozzle can be compensated by adjusting the value of $\na$ in our model. Bending stiffness is less important for thinner jets, and surface tension for larger flow velocities. For $\vb$ close to zero, the jet is unstable near the belt, but adequate modeling of this instability is still an open question.

To conclude, we state that our model predicts correctly the transitions between the parameter regions for the three flow regimes. Also the tendencies in the (partial) monotonic behavior of $\xe$ as $\vb$ increases are predicted well, yielding a satisfactory qualitative agreement. However, significant quantitative differences are obtained.
\section{Summary of the three flow regimes} \label{sec:SummaryThreeFR}
Using our knowledge about the three flow regimes from model and experiment, we describe typical features of each flow regime. In the model the three flow regimes are characterized by the sign of the dimensionless variable $\xi$. The value of $\xi$  represents the momentum transfer through a cross-section of the jet and describes the balance between the inertia and viscous terms in the conservation of momentum equation (\ref{eq:stjet_cons_mom1}). Flow characterization using experimental jet shape features is possible as well. Below, we describe each flow regime separately
\begin{itemize}
\item[\textbf{Concave flow.}] In this flow regime $\xi$ is positive. This means that the momentum transfer due to inertia is larger than that due to viscosity. This is reflected in the concave shape of the jet comparable to a ballistic trajectory. The nozzle orientation is important for the jet shape. When the nozzle points vertically down the jet shape in this flow regime is vertical, no matter the flow regime is concave or vertical. Therefore, in this case the characterization of the flow regime using the jet shape does not distinguish between vertical and concave jets.
\item[\textbf{Vertical flow.}] In this flow regime $\xi$ changes sign from negative near the nozzle to positive near the belt. Hence, the momentum transfer due to viscosity is larger near the nozzle and the one due to inertia is so near the belt. The belt and nozzle orientations are now irrelevant for the jet shape, which is straight vertical in the experiments (except a possible bending region near the nozzle and bending or unstable region near the belt) as well as in the model.
\item[\textbf{Convex flow.}] In this flow regime $\xi$ is negative, which means that the momentum transfer due to viscosity is larger than that due to inertia. Both in the experiments and the model the jet shape is convex (disregarding a small bending region near the nozzle in the experiment) and the jet touches the belt tangentially .
\end{itemize}
Summarizing, we conclude that the flow regimes can be characterized by the sign of the momentum transfer through the cross-section of the jet or by the convexity of the jet shape. However, for $\na=\pi/2$ the concave jet shape is vertical, which makes it then impossible to distinguish between the concave and vertical flow regimes. Some more shape features such as the tangency condition at the belt for the convex flow, and the relevance of the nozzle orientation for the concave flow can be used to distinguish these flow regimes.
\section{Conclusions} \label{sec:Conclusions}

In this paper we have studied experimentally and theoretically the problem of the fall of a viscous jet onto a moving belt. Three flow regimes of the jet are distinguished and characterized by the convexity of the jet shape, i.e. concave, vertical, and convex.

We have modeled the jet using a thin-jet approximation including the effects of inertia, viscous tension and gravity. The model consists of the stationary conservation laws for mass and momentum. A change of the independent variable is made to allow for a transformation of the model equations into an algebraic equation. The partitioning of the parameter space between the three flow regimes is evaluated in terms of three dimensionless numbers.

The model shows that the sign of the momentum transfer through a cross-section of the jet determines the corresponding flow regime. For each flow regime the correct boundary condition for the jet orientation is derived by looking at the characteristics of the dynamic conservation of momentum equation.
These boundary conditions for the jet orientation are:
\begin{enumerate}
 \item the nozzle orientation for the concave jet,
 \item no boundary condition for the vertical jet,
 \item the tangency of the jet at the belt for the convex jet.
\end{enumerate}
The missing boundary condition for the vertical jet is replaced by the constraint that at the point where the momentum transfer equals zero the jet is aligned with the vertical direction of gravity.

It is shown that a continuous transition between the concave and the convex jets is only possible via the vertical one. Also  the way how the dimensionfull parameters should be changed in order to leave the convex or concave jet region is indicated.

Comparison of the relations between the horizontal position of the touchdown point $\xe$ and the belt velocity $\vb$, obtained from experiments and from the model, shows that:
\begin{enumerate}
 \item The model and experiments show similar monotonic behavior of $\xe$ as $\vb$ is changed.
 \item The parameter regions in the $(\vb,L)$-plane for the three flow regimes predicted by the model agree with the experimental data.
 \item Quantitatively the relations between $\xe$ and $\vb$ show a significant mismatch (experiments give smaller $\xe$) due to differences in the shapes of calculated and experimentally observed jets.
\end{enumerate}
As a final conclusion, we state that the model, which includes viscous tension and inertia, but disregards air resistance, bending stiffness, and surface tension, describes in qualitative sense the fall of a jet of a Newtonian fluid under gravity.
\section{Acknowledgments}
\begin{acknowledgments}
The authors would like to acknowledge Teijin Aramid, a part of the Teijin group of companies, and especially Hans Meerman for providing the experimental equipment and valuable suggestions for experiments.
\end{acknowledgments}
\end{document}